\documentclass{aa}  

\usepackage{graphicx}
\usepackage{txfonts}
\usepackage[colorlinks=true]{hyperref}
\usepackage{natbib}
\bibpunct{(}{)}{;}{a}{}{,}
\newcommand{\um}{$\mu$m}

\def\utw{\smash{\rlap{\lower5pt\hbox{$\sim$}}}}
\def\udtw{\smash{\rlap{\lower6pt\hbox{$\approx$}}}}

\def\Teff{\hbox{\it T$_{\rm eff}$}}

\def\Mbol{\hbox{\it M$_{bol}$}}
\def\mbol{\hbox{\it m$_{bol}$}}

\def\Teff{\hbox{\it T$_{\rm eff}$}}

\def\Ql{\hbox{\it Q$_{\lambda}$}}

\newcommand{\Ks}{{\it K$_{\rm s}$}}

\newcommand{\Al}{{\it A$_\lambda$}}
\newcommand{\Aks}{{\it A$_{\it K_{\rm s}}$}}
\newcommand{\Ak}{{\it A$_{\it K}$}}
\newcommand{\Av}{{\it A$_{\it V}$}}
\newcommand{\Akint}{{\it A$_{\it K_{\rm s}}{\rm (int)}$}}
\newcommand{\Akenv}{{\it A$_{\it K_{\rm s}}{\rm (env)}$}}
\newcommand{\Aktot}{{\it A$_{\it K_{\rm s}}{\rm (tot)}$}}

\def\BCl{\hbox{\it BC$_{\lambda}$}}

\def\BCKs{\hbox{\it BC$_{\it K_S}$}}

\def\simgr{\mathrel{\hbox{\rlap{\hbox{\lower4pt\hbox{$\sim$}}}\hbox{$>$}}}}

\begin{document}

   \title{Infrared colours and bolometric corrections
   of SiO masing stars in the inner Milky Way }


\author{Maria Messineo
          \inst{1,2}\fnmsep\thanks{This work was partly carried out
over the past year, while MM was a freelancer.}
           }

   \institute{Dipartimento di Fisica e Astronomia 
“Augusto Righi”, Alma Mater Studiorum, Università di Bologna, Via Gobetti 93/2, 
I-40129 Bologna, Italy\\
              \email{maria.messineo2@unibo.it}
         \and
             INAF - Osservatorio di Astrofisica e Scienza dello Spazio 
di Bologna, Via Gobetti 93/3, I-40129 Bologna, Italy\\
             }

   \date{Received November 30, 2022; accepted xx xx, xxxx}

 
  \abstract
   {} 
  {We analysed a sample of SiO-masing late-type stars located in the inner Galaxy with the goal of setting better constraints on their obscuration.  }
  { This reference sample
has allowed us to define mathematical relations between their
dereddened infrared colours and the observed colours 
(e.g. \Ks-[8], \Ks-[24]).
} 
  {The derived equations define a property (the locus) of these
late-type stars. Therefore, they enable us to
derive the interstellar extinction.
With estimated spectral types, it is possible to 
decompose the total extinction in 
the two components (interstellar and envelope extinction).
 }
  {These relations are useful for classifying extremely 
obscured late-type stars
located in the inner Galaxy. 
Estimating  the two extinction components can be performed 
on an individual late-type star, 
independently of its surroundings and even when   few
mid-infrared measurements are available.
}


   \maketitle
%

\section{Introduction}

Typically, SiO masing stars have late spectral types, namely, later than M4-M5.
Dusty circumstellar envelopes surround the central star,
where  maser emission from  SiO, OH, and H$_2$O molecules may originate.
The starlight is absorbed, reprocessed, and re-emitted 
at infrared wavelengths 
by the dusty envelope.
In other words, the envelope represents a deforming and transforming mirror of the original 
stellar energy distribution (SED).
The effect is dependent on both the interstellar absorption 
caused by a diffuse medium 
along the line of sight and on the optical depth of the envelope, 
which can range from zero to an equivalent of about \Av = 100 mag. 
In the inner Galaxy, the interstellar extinction can likewise 
reach \Av = 30–40 mag. 
Due to the severe dust obscuration and the 
patchy interstellar extinction, 
the reddening,  distances, and luminosities of late-type 
stars cannot  easily be inferred.

To characterize the SED of an asymptotic giant  branch (AGB) 
star, a decomposition of its total extinction into
the interstellar component and the circumstellar component
is required. 
\citet{messineo05} measured the total extinction (\Aks)  
for a sample of masing AGB stars and broke it down into \Akint\ 
for the interstellar medium (ISM) and \Akenv\ for the envelope.
Their method was straightforward and relied on multi-wavelength 
(single-epoch) photometric measurements of the targeted AGB star,  
covering the near- and mid-infrared window.
They  assumed that the median extinction of nearby giants 
could be used to approximate the interstellar extinction.

Stellar colours and bolometric corrections (BC) of an AGB star
vary depending on the properties of its envelope, namely, the extent,
density,  and mass loss rate.
Here, we present correlations between stellar colours,  
bolometric corrections, and envelope types obtained 
for the sample of SiO masing stars from \citet{messineo18}.
The advantage of this sample is that it covers  a wide range
of interstellar extinction and infrared stellar colours. 
We used it to elaborate
on the intrinsic colours of cold stars
in \citet{messineothesis}, \citet{messineo05}, 
and \citet{messineo12}.
This time, we have used this reference sample 
to develop a new technique. 
We have established correlations between the intrinsic infrared colours
and the observed infrared colours and envelope types, assuming
that we  know the Galactic interstellar absorption curve. 
By solving a system of equations, 
it is then technically possible to decompose the extinction into 
the interstellar and envelope components, 
independently of the star's surroundings. 
It is possible to model the BCs as a function of the 
intrinsic stellar colours 
and to determine the apparent bolometric magnitudes.

This idea is especially helpful  in the innermost obscured regions 
of the Milky Way, 
where it is not possible to study the entire SED, 
and individual measurements  in the mid-infrared are often 
of poor quality (e.g., because of confusion).
If the nature of the star is known, for example,
because of its maser emission,
we can use the established equations
for the \Ks-[24] or \Ks-[8] colours to determine
the  extinction,  characterize  the  SED of the star, 
and  estimate the apparent bolometric magnitude.

We describe the stellar sample 
and the available parameters in Sect. \ref{sec.catalog}.
Extinction-free colours are defined and the relationships between them 
and the dereddened colours are described in Sect. \ref{Ql}. 
The existence of these 
relations allows us to determine the  interstellar extinction.
Using an average spectral type (such as M6), 
we can also estimate the envelope extinction given the 
interstellar extinction.
Synthetic infrared colours derived from the DUSTY models are 
shown and compared with the observational data in 
Sect. \ref{grid-black}. The empirical estimates of extinction 
are compared with previously 
published estimates in Sect. \ref{verifica_maser}.

\section{ Sample }
\label{sec.catalog}

The sample consists of 572   late-type stars
in the process of losing mass, which we searched to find SiO maser emission 
\citep{messineo02,messineo18}.  The sources are located
at low latitudes ($|b|<0.5^\circ$).
The sample had been created  using near- and mid-infrared
photometric measurements. The collected  measurements  
cover from 0.9 \um\ to 24 \um\ 
and come from the publicly available surveys
2MASS, MSX, WISE, and GLIMPSE\footnote{
2MASS stands for Two Micron All Sky Survey
\citep{skrutskie06}, DENIS for Deep Near Infrared Survey of the Southern
Sky \citep{epchtein94},  MSX for Midcourse Space Experiment 
\citep{price01},  WISE for Wide-field Infrared
Survey Explorer \citep{wright10}, and
GLIMPSE for Galactic Legacy Infrared Mid-Plane Survey
Extraordinaire \citep{churchwell09}.}   \citep{messineo18}.
Each photometric data point is from a single epoch.
Several stellar parameters were estimated by \citet{messineo05} and
\citet{messineo18}.

\noindent
{\it Total \Aks\ determination.}
The total extinction was estimated by \citet[][]{messineo05,messineo18}
in the \Ks\  versus $J-$\Ks\ or $H-$\Ks\  diagram, 
by shifting the SiO masing stars onto the
47 Tuc giant branch along the reddening vector -- 
which implies a late M type (M4-M5) star. 
The  values  of \Aktot\ span from 0 to 5.5 mag.

\noindent
{\it Interstellar \Aks\ determination.}
The mean of surrounding stars is used as a proxy 
for \Akint\ \citep{messineo05,messineo18}. 
This assumption relies on the strong hypotheses 
that red giants and AGBs share the same Galactic distribution
and that the target star is situated at the peak of the stellar 
density along the line of sight.
Masing AGB stars are good tracers of the thin central bar, 
as shown by \cite{habing06}. 
The thin disk population and the thick disk population do, 
in fact, cohabit in the disk; however, at low latitudes, 
the thick disk's contribution is less than 10\% \citep[][]{lee11}.  

\noindent
{\it Apparent bolometric magnitudes.}
\citet{messineo18} computed
apparent bolometric magnitudes
by directly integrating under 
the stellar SEDs from 0.9 \um\ to 24 \um. 
At longer wavelengths, the flux was linearly extrapolated to zero.   
At short wavelengths, the flux was estimated with 
a blackbody of 3000 K.

\noindent
{\it Infrared bolometric corrections (BC).}
For a given photometric band,  the BC 
represents the correction to be added  to the
dereddened apparent magnitude in order
to estimate the stellar apparent bolometric magnitude, (\mbol),
$BC_\lambda$=\mbol-m$_\lambda$. 
There is a tight relation between the 
infrared \BCKs\ values and the intrinsic stellar colours of \citet{messineo18}.
Second-order polynomial  fits to the data are
given in the appendix.

\noindent
{\it Large-amplitude variables.}
The sample from \citet[][]{messineo05}  is made up of Mira-like stars.
The average photometric measurements should ideally be used to infer 
average colours and luminosities due to 
large periodic light variations.
Amplitudes of Miras in the \Ks\ band can reach up to 2 mag, whereas
at longer wavelengths, they are substantially smaller
(e.g. at 15 \um\ they are about 50\% than those in \Ks).
Nevertheless, average quantities of the entire sample generate quantities
typical of the actual average value \citep{messineo04},
based on the random phase of each star 
(absence of synchronous pulsations).
Using DENIS, 2MASS, MSX, and WISE photometric flags,
\citet{messineo18} found that at least 74\% 
of the sample is made of long-period variables (LPV).

There are 282 out of 572 stars with matches in the Gaia DR3 main catalog \citep{gaia23},
249  have a measured G-band magnitude; of those, 72\%  (179) 
are classified as
variables; the G-band values range from 9 to 20 mag;  
there are from 10 to 62
G-band observations per star, from 2 to 60 BP-band observations, and
from  7 to  65 RP-observations.
So far, Gaia DR3 has released photometric data covering 
34 months of observations \citep{eyer23}.
Two scanning laws were used; the ecliptic pole
scanning law (EPSL) during the first 28 days 
and the nominal scanning law (NSL) afterward;
the observational cadence depends on the period of the spin axis (6 h) 
and  the precession around the Sun (63 d); 
the scanning law favours the Ecliptic poles.
At the end of the six-year mission, each star is supposed to have
an average of $\approx 70$ usable photometric transits.

For the targets  recognised as Gaia variables, 
the differences between the highest and lowest G-band  measurements
range from 0.12 mag to 3.54 mag \citep{eyer23}.
Those marked as `NOT\_AVAILABLE' regarding variability
are fainter (mostly from G=18 to 21 mag), with a peak at G=20 mag
and have from 0 to 43
G-band observations, from 0 to 19 BP-band observations, and
from  0 to  38 RP-observations.
The differences between the highest and lowest G-band magnitudes 
range from 0.0 mag to 0.67 mag.
A fraction of 55\% of  Gaia matches are variables in Gaia
and also in \citet{messineo18} based on infrared measurements.
Gaia adds  16\% of new variables (not detected  via  infrared flags).

The amplitude, or the difference between the minimum and maximum, 
is typically more than G=0.8 mag for Mira AGBs 
\citep[e.g.][]{lebzelter23,messineo22}.
With variations greater than 0.8 mag, 
74\% of the Gaia variables are most likely Miras.
However, only 86 of these 179 Gaia variables --
whose amplitudes and periods range from 0.12 mag to 3.34 mag and 
their periods from 42 d to 808 d --
are currently listed in the Gaia catalog of classified 
LPVs \citep[][]{lebzelter23}.
The Gaia  catalog of    LPVs includes only
stars with a 5-95\% quantile range larger than 0.1 mag in the G-band.

Since there is no discernible difference between the 
population of Gaia-identified LPVs and those not 
designated as variables for the purposes and diagrams 
displayed in this study, 
the fits and diagrams presented in this paper employ 
the whole sample of stars in \cite{messineo18}.

\begin{figure*}
\begin{center}
\resizebox{0.33\hsize}{!}{\includegraphics[angle=0]{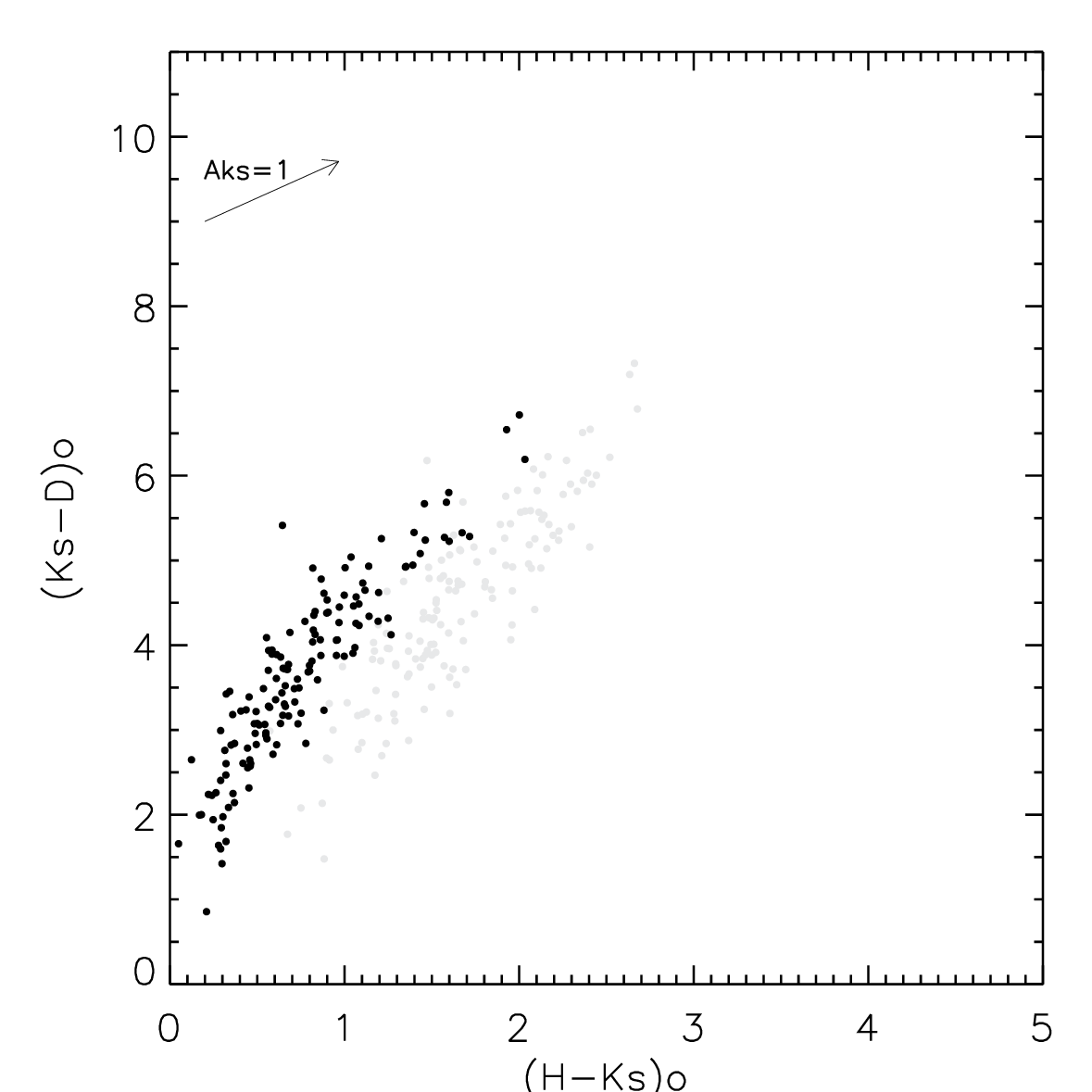}}
\resizebox{0.33\hsize}{!}{\includegraphics[angle=0]{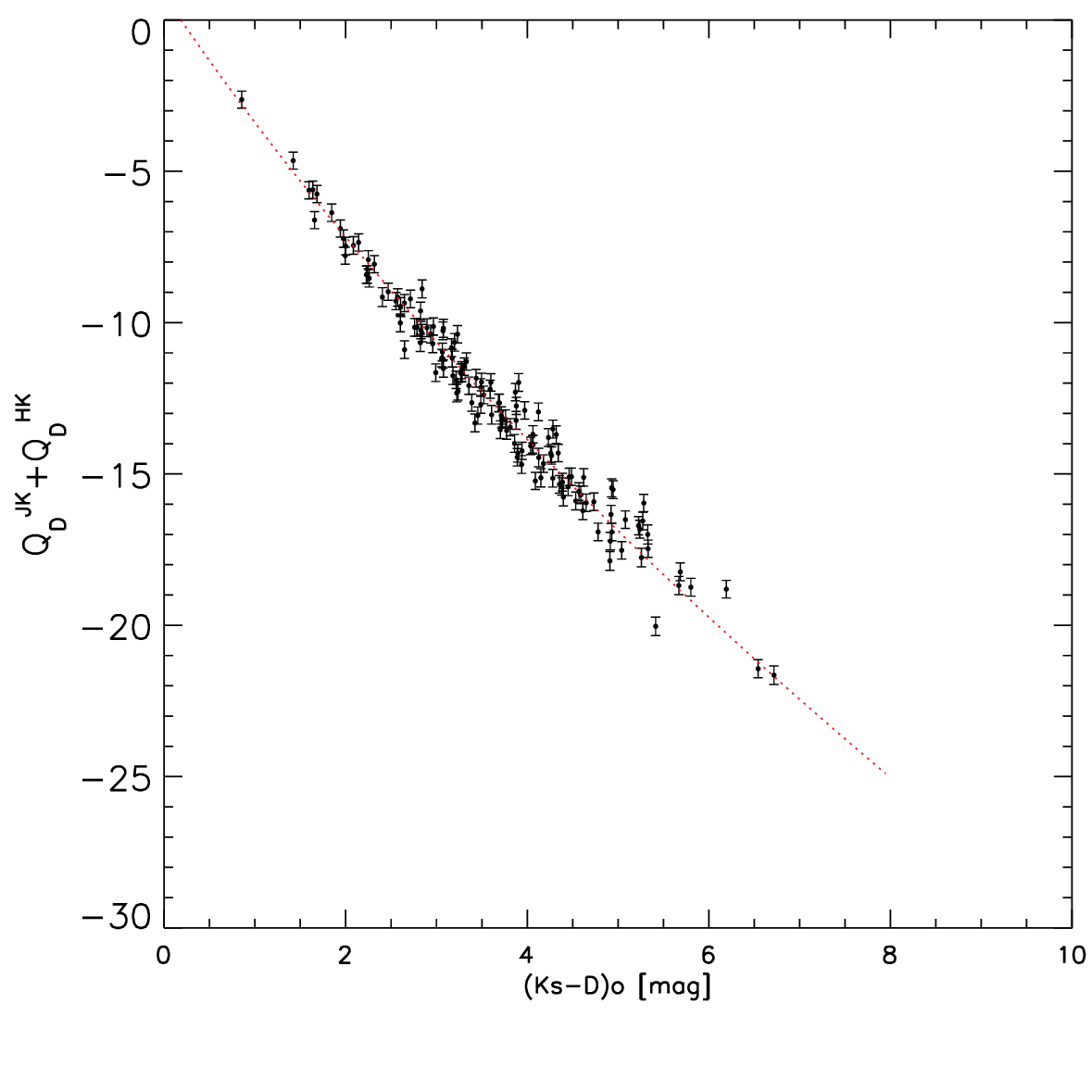}}
\resizebox{0.33\hsize}{!}{\includegraphics[angle=0]{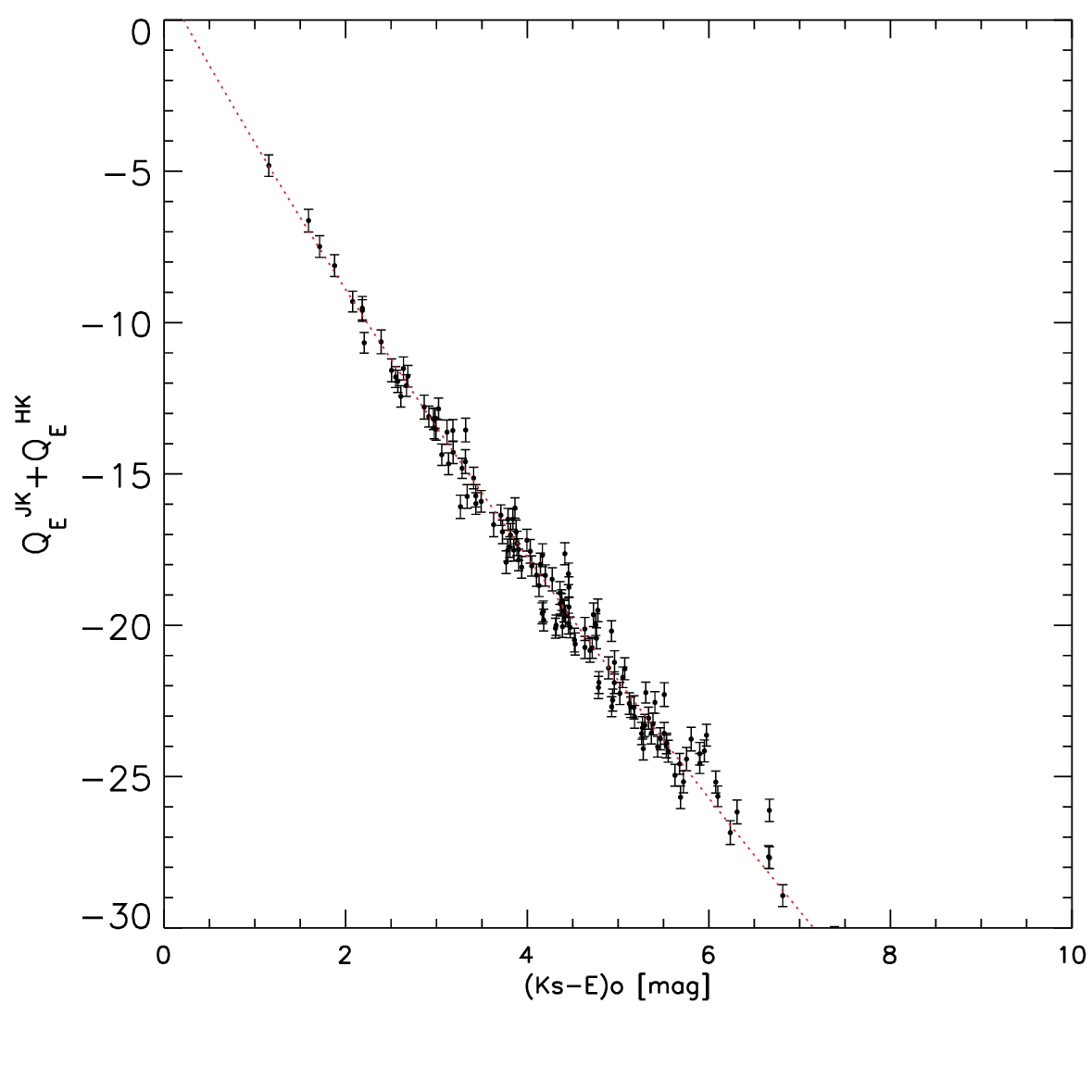}}
\end{center}
\caption{\label{fig.bcD}  
 Correlations between intrinsec colours and between Q($\lambda$) values
and intrinsec colours. 
{\it Left panel:}
Intrinsic (\Ks$-[D]$)o versus ($H-$\Ks)o colours
of \citet{messineo05}, de-reddened using the median interstellar extinction 
of surrounding stars, shown in black. 
For comparison, in gray, we give the observed (\Ks$-[D]$) versus ($H-$\Ks) colours.
{\it Middle panel:} Q(D) 
colours vs. (\Ks$-[D]$)o, for the MSX D-band.  
{\it  Right panel:} Q(E) 
colours versus (\Ks$-[E]$)o, for the MSX E-band.  The red dotted lines 
are  fits to the data points. Error bars on the Q(D) and Q(E)
are obtained by propagating the errors on the $J$, $H$, \Ks,  $D$, and $E$ magnitudes.
}
\end{figure*}


\section{ Extinction}
\label{Ql}

\citet{messineo12} used
the interstellar extinction-free parameters $Q1$ and $Q2$, where\\
$Q1 = (J-H)-1.8 \times (H-K_{\rm s})$, and $Q2 = (J-K_{\rm s}) - 2.69 \times (K_{\rm s}-[8.0])$\footnote{The 2.69 coefficient
is obtained using the effective extinction values of \citet{indebetouw05}.}
to study the properties of different types of late-type stars.
 The extinction-free colours can be calculated directly from the observations.\\
\citet{messineo12} showed that the intrinsic MSX \Ks$-[A]$ colours, 
(\Ks$-[A])_o$, of SiO masing stars  are well correlated 
with the $Q1$ parameter, as well as 
with their BC($[A]$) values.
The $Q1$ parameter spans a quite narrow range (1.5 mag), while
the $Q2$ parameter spans more than 10 mag (from 0 to 12 mag) and 
is a very sensitive meter of circumstellar obscuration.

This correlation is of great advantage because it allows us to 
decompose the total \Aks\ in the envelope component, \Aks(env), and 
the interstellar component, \Aks(int), 
and to obtain  estimates of the stellar bolometric magnitudes.
In this exercise, we revisit this idea
to determine    \Aks(int) and  \Aks(env) quantitatively
by taking the following steps.

By having  adopted an interstellar  extinction curve,
we determine the \Ql\ values for stars in \citet{messineo05}
that have  \Aks(int) values.
Then, we verify the existence of mathematical relations 
between the \Ql\ values
and independently obtained, intrinsic colours.
For stars with unknown \Aks(int), 
we use these relations plus the observed \Ql\ values to estimate the
intrinsic colours and \Aks(int).
By using \Aktot\ and \Ak(int), we also  estimate
the circumstellar extinction, \Aks(env).
Eventually, we obtain some constraints on the circumstellar extinction 
curve, as suggested by the kind referee.

\subsection{ Adopted interstellar extinction ratios}

\subsubsection{ NIR extinction power law}

At near-infrared (NIR) wavelengths, the Galactic extinction curve is usually 
approximated with a power law, {\it \Al} $\propto \lambda^{-\alpha},$
and it appears the same in every direction. 
$\alpha$ determinations in the literature range from 1.61 to 2.2. 
A value of  $\alpha = 1.61$ was measured by \citet{rieke85}.
A model with $\alpha=1.9$ was favored in \cite{messineo05} 
based on an analysis of the 2MASS colours of late-type stars 
in the inner Galaxy. 
$\alpha$ measurements  larger than 2.0 were published 
in a number of works between 2006 and 2011 
\citep[e.g.,][]{nishiyama06,stead09}. In the nuclear disk,
\citet{fritz11} reported a value  of 2.1.
For late-type stars at the Galactic center, 
which suffer about 2 mag of reddening in bands $H-$\Ks, 
the increase of $\alpha$ from  1.61 to  1.9 
implies a decrease of \Aks\ of 0.7 mag. 
A change in $\alpha$ from   1.9 to  2.1 
implies another decrease of about 0.35 mag. A higher $\alpha$ 
decreases the extinction and makes the stars fainter.

In \citet{messineo05}, an $\alpha$ of 1.9 was adopted,  
\Aks/E(J-Ks)=0.537 and \Aks/E(H-Ks)=1.493, and,
$A_J$:$A_H$:$A_{Ks}$=2.86:1.66:1.00.
Using the same calculation performed  in \citet{messineo05},
for $\alpha$=2.1, it is found that
\Aks/E(J-Ks)=0.455 and \Aks/E(H-Ks)=1.311, and,
$A_J$:$A_H$:$A_{Ks}$=3.208:1.766:1.000.
Within errors, the latter ratios are  consistent with the 
empirical measurements of \citet{wang19}:
$A_J$:$A_H$:$A_{Ks}$=3.115:1.679:1.000.
\Aks/E(J-Ks)=0.473 and \Aks/E(H-Ks)=1.472.
At the Galactic center, \citet{nogueras20}
measured $A_J/A_H$ = 1.87 $\pm$ 0.03 and $A_H /A_K$ = 1.84 $\pm$ 0.03,
the power law with $\alpha$=$2.1$ yields
$A_J/A_H$ = 1.82        and $A_H/A_{Ks}$ = 1.77;
while the Wang's ratios yield
$A_J/A_H$ = 1.85        and $A_H/A_{Ks}$ = 1.68.
Here, a single  power law with index $-\alpha$= $-2.1$ is adopted.

\subsubsection{ MIR extinction  law}

The average effective extinction values for the filters of the MSX, WISE, 
and GLIMPSE surveys were estimated by convolving the extinction curves 
with a 3200 K giant of \citet{allard11} and  the filter response curves.

As in \citet{messineo05}, the parametric extinction curve of 
\citet[][]{rosenthal00}, denoted as "curve 2,"
obtained in the OMC-1 cloud with H$_2$ observations, was used
along with two other curves.
Curve 3 is  a modified version of  the Rosenthal curve  in
the 3-8 \um\ region to incorporate the ratios from H-line observations 
in the Galactic center region of \citet{fritz11}.\footnote{ This is
a revised version of curve 3 in \citet{messineo05}, where
the 3-8 \um\ opacity enhancement  had been taken as a constant.}
Curve 4 was derived  by \citet{gordon21}  with Spitzer 
spectroscopic data of early-type stars
located in the direction of a diffuse medium. 
The stars used by Gordon et al. are located at A$_{V}$ $<$ 5 mag.
The average  effective extinction values 
from curve 2, curve 3, and curve 4 are listed 
in Table \ref{table:extinction}.

The effective extinction ratios may vary at mid-infrared wavelengths.
The peak of the silicate feature around 8-9 \um\
$\frac{A_{9.7}}{ A_{\it K_{\rm s}}}$ is 1-1.3 in the Galactic center, 
but it is  only 0.7 in 
the solar neighbouring stars observed by Spitzer.
The Gordon curve can be regarded as a lower limit curve, while
the Galactic center curve represents  an upper limit curve.

The effective extinction ratios for the WISE and GLIMPSE filters 
were also determined by \citet{wang19}  using broad-band 
photometry of red clump stars, 
which are primary indicators of distances due to their  
known luminosity and colours.
These are also shown  in the  Table \ref{table:extinction}
and seem to agree   with our predictions from the curves.

\begin{table*}
\renewcommand{\arraystretch}{0.6}
\caption{\label{table:extinction} MIR  interstellar effective extinction,
$\langle$A$\rangle$/A$_{K_S}$.
}
\begin{centering}
\begin{tabular}{lrrccccccc}
\hline
\hline
{\rm Filter}& $\lambda_{\rm ref}$&$\Delta \lambda$ &Curve 2 (Ros)$^*$ & Curve 3 (Ros+Fritz)$^{**}$ & Curve 4 (Gordon)$^{***}$ & Wang$^{****}$ & C$_\lambda$ (2) & C$_\lambda$ (3) & C$_\lambda$ (4)\\
&&& ($A_{9.7}/A_{2.12}=1.0$)& ($A_{9.7}/A_{2.12}=1.0$)& ($A_{9.7}/A_{2.12}\approx0.7$)\\
&&& {\rm $\langle$A$\rangle$/A$_{K_S}$} &{\rm $\langle$A$\rangle$/A$_{K_S}$}&{\rm $\langle$A$\rangle$/A$_{K_S}$}&{\rm $\langle$A$\rangle$/A$_{K_S}$}\\
{\rm }&[{\rm $\mu$m }]& [{\rm  $\mu$m}]&\\
\hline
{\rm MSX A}         &8.28&4.0  & 0.380&0.490&0.340&             &4.797&  5.831 & 4.506\\
{\rm MSX C}         &12.1&2.1  & 0.491&0.491&0.320&             &5.843&  5.843 & 4.374\\
{\rm MSX D}         &14.6&2.4  & 0.292&0.292&0.219&             &4.201&  4.201 & 3.808\\
{\rm MSX E}         &21.3&6.9  & 0.405&0.405&0.300&             &4.998&  4.998 & 4.249\\
{\rm Glimpse [3.6]} &3.55&0.75 & 0.435&0.447&0.486&$0.47\pm0.04$&5.264&  5.378 & 5.786\\
{\rm Glimpse [4.5]} &4.49&1.01 & 0.298&0.402&0.351&$0.33\pm0.04$&4.237&  4.973 & 4.582\\
{\rm Glimpse [5.8]} &5.73&1.42 & 0.196&0.368&0.250&$0.24\pm0.04$&3.699&  4.706 & 3.965\\
{\rm Glimpse [8.0]} &7.87&2.93 & 0.295&0.408&0.266&$0.32\pm0.04$&4.218&  5.024 & 4.052\\
{\rm WISE W1}       &3.35&0.66 & 0.475&0.479&0.525&$0.50\pm0.05$&5.665&  5.708 & 6.261\\
{\rm WISE W2}       &4.60&1.04 & 0.284&0.398&0.337&$0.33\pm0.05$&4.154&  4.940 & 4.486\\
{\rm WISE W3}       &11.56&5.51& 0.527&0.534&0.376&$0.51\pm0.12$&6.288&  6.382 & 4.766\\
{\rm WISE W4}       &22.09&4.10& 0.394&0.394&0.294&             &4.908&  4.908 & 4.213\\
\hline
\end{tabular}
\end{centering}
\begin{list}{}
\item WISE filter specifications are from \citet{jarrett11}.
\item SPITZER filter specifications are from \citet{fazio04}.
\item MSX filter wavelengths are from 
${{\rm https://irsa.ipac.caltech.edu/Missions/msx.html.}}$
\item ($^*$) This column is based on the curve 2 of \citet{messineo05},
which is the \citet[][]{rosenthal00} curve.
\item ($^{**}$) curve 3 is again the \citet[][]{rosenthal00} curve 
modified  in the at 3-8 \um\ to include
the more recent observations of \citet{fritz11}.
\item ($^{***}$) This column is based on the interstellar extinction
curve obtained by \citet{gordon21}
using stellar spectra taken with the Spitzer satellite.
\item ($^{****}$) Values obtained with broad-band photometry by \citet{wang19}.
\item The C$_\lambda$=C$_\lambda^{JKs}$+C$_\lambda^{HKs}$ 
values are calculated using Eq. (2) and Eq. (4).
C$_\lambda$(2) refers to curve 2, C$_\lambda$(3) refers to curve 3, and
C$_\lambda$(4) refers to curve 4.
\end{list}
\end{table*}

\subsection{ Definitions of the \Ql\ parameters}

Based on parameters $Q1$ and $Q2$ of \citet{messineo12},   
we can analogously calculate 
an interstellar extinction-free parameter \Ql\ 
for each filter and each
 interstellar extinction curve (Table \ref{table:extinction}):
\begin{equation}
Q_\lambda^{JKs}=(J-Ks)-C^{JKs}_\lambda \times (Ks-[\lambda]),
\end{equation}
where 
\begin{equation}
C^{JKs}_\lambda=\frac{\frac{A_{ J}}{A_{ K_{\rm s}}}-1.}{1.-\frac{A_{\lambda}}{A_{ K_{\rm s}}}}
\end{equation}

and 

\begin{equation}
Q_\lambda^{HKs}=(H-Ks)-C^{HKs}_\lambda \times (Ks-[\lambda]),
\end{equation}
where 
\begin{equation}
C^{HKs}_\lambda=\frac{\frac{A_{ H}}{A_{ K_{\rm s}}}-1.}{1.-\frac{A_{\lambda}}{A_{ K_{\rm s}}}}
.\end{equation}

By combining Eqs. 1 and 3, we obtain:
\begin{equation}
Q_\lambda=Q_\lambda^{HKs}+Q_\lambda^{JKs}=(J+H-2Ks)-(C^{JKs}_\lambda+C^{HKs}_\lambda) \times (Ks-[\lambda]).
\end{equation}

The constants can be estimated with an extinction curve.
Curve 3 applies to highly obscured regions.
The Gordon's curve (for a diffuse interstellar gas)
can be regarded as the Galactic lower limit (smaller peak at 10 \um). 

The filters GLIMPSE [8], WISE W3, and MSX A contain the 10 \um\ 
silicate emission. The  MSX D and WISE W4 filters
appear to be particularly useful as they are not  
affected by the silicate emission 
and show small variations of  $C_\lambda$ values 
in the three curves.

By definition, Eq. 5 yields an identical value when  
calculated with the
observed magnitudes and  the dereddened 
magnitudes (see Appendix). 
This allows us to estimate the intrinsic colours.

 In principle, we can use separately equations (1) and (3)
and obtain two independent colour estimates. However, the $Q_\lambda^{HKs}$ values 
span a much smaller range (e.g., $Q_D^{HKs}$ from $-1$ to $-5$ mag)
than the $Q_\lambda^{JKs}$ values (e.g., $Q_D^{JKs}$ from $-2$ to $-19$ mag) 
and the interpolated colours  have large errors.
The combined $Q_\lambda=Q_\lambda^{HKs}+Q_\lambda^{JKs}$ colours span a wider  range, 
for instance,\ $Q_D$ ranges from $-2$ to $-22$ mag (see Fig. \ref{fig.bcD}).
The combined case is used in the following sections.

\subsection{\Ql-intrinsic-colours curves for stars with known  \Aks(int)}
Using the stars   by \citet{messineo18} as a reference sample, 
a series of equations is set up to express the intrinsic colours
as a function of the $Q_\lambda$ parameters.
The parameter \Ql\  is correlated with the dereddened 
$(Ks-[\lambda])$ colour, for instance, $Q(D) \propto (Ks-D)o$ 
(see Fig.\ \ref{fig.bcD}).
The fits are made with third-grade polynomial fits.
It follows that:  
\begin{equation}\label{equ1}
(Ks-\lambda)_o= {\rm c1} \times Q(\lambda)+ {\rm c2} \times Q(\lambda)^2+
 {\rm c3} \times Q(\lambda)^3 + {\rm c0.} 
\end{equation}
The  intrinsic colours are  calculated with the median \Akint\  
values of  the surrounding stars \citep{messineo05}, which are 
rescaled to a power law
with an index of $-2.1$, and the assumed
effective extinction ratios.
In \citet{messineo04} and \citet{messineo05}, 
it is shown that the  SiO masing stars
are rare and the majority of the surrounding stars
are normal giants. 
The large interval spanned by \Ql\ is due to the 
intrinsic colour range  of the stars (stellar SEDs).
These bright late-type stars  are mostly
 mass-losing Mira AGBs with spectral types later than M4.

For each filter, the constants of the fits
are listed in Table \ref{table.qxfits},  when using the  
revised curve 3 of \citet{messineo05} and in Table 
\ref{table.qxfits.gordon} for the extinction curve of \citet{gordon21}.

\begin{table*}
\caption{\label{table.qxfits} Coefficients of the polynomial fits between 
(\Ks$-[\lambda]$)o colours and the \Ql\ parameters, 
using curve 3.}
\begin{tabular}{llrrrrr}
\hline
\hline
      X-axis &         Y-axis &    Coef n1               & Coef n2           &  Coef n3            & Coef n0             & $\sigma$  \\
\hline
        QA    &   (Ks$-$A)o    &  $-$0.1825$\pm$0.0319   & 0.0006$\pm$0.0028 & 0.0000$\pm$0.0001   & 0.1686$\pm$0.1108    &  0.15\\
        QC    &   (Ks$-$C)o    &  $-$0.1453$\pm$0.0456   & 0.0021$\pm$0.0026 & 0.0000$\pm$0.0000   & 0.2636$\pm$0.2535    &  0.15\\
        QD    &   (Ks$-$D)o    &  $-$0.2296$\pm$0.0261   & 0.0033$\pm$0.0010 &   $..$              & 0.1846$\pm$0.1619    &  0.20\\
        QE    &   (Ks$-$E)o    &  $-$0.1875$\pm$0.0162   & 0.0015$\pm$0.0004 &   $..$              & 0.2133$\pm$0.1474    &  0.18\\

\hline
       QW3    &  (Ks$-$W3)o    & $-$0.1355$\pm$0.0162    & 0.0026$\pm$0.0009 &  0.0004$\pm$0.0000  & 0.2200$\pm$0.0937    & 0.15\\
       QW4    &  (Ks$-$W4)o    & $-$0.0887$\pm$0.0374    & 0.0081$\pm$0.0020 &  0.0001$\pm$0.0000  & 0.7127$\pm$0.2212    & 0.18\\

\hline
       Q58    &(Ks$$-$[5.8]$)o  & $-$0.2853$\pm$0.0377   &$-$0.0013$\pm$0.0075&$-$0.0000$\pm$0.0004&0.2301$\pm$0.0519 &0.20 \\
       Q80    &(Ks$$-$[8.0]$)o  & $-$0.2351$\pm$0.0331   &   0.0018$\pm$0.0047& 0.0000$\pm$0.0002  &0.1736$\pm$0.0669 &0.18  \\

\hline
\end{tabular}
\begin{list}{}
  \item  {\bf Note:} Y-Fit = Coef~n1 $\times$ X-axis  + Coef~n2 $\times$ (X-axis)$^2$ + Coef~n3 $\times$ (X-axis)$^3$ +Coef~n0. \\
  The colours  used for the fit are those dereddened with the \Akint\ from surrounding stars
  by \citet{messineo05} and \citet{messineo18} (rescaled to a power law with an index of $-2.1$)
  and effective ratios from curve 3 in Table \ref{table:extinction}.
\end{list}
\end{table*}

\begin{table*}
\caption{\label{table.qxfits.gordon}  Coefficients of the polynomial fits between 
(\Ks$-[\lambda]$)o colours and the \Ql\ parameters,
based on the interstellar extinction curve by \citet{gordon21}.} 
\begin{tabular}{llrrrrr}
\hline
\hline
      X-axis &         Y-axis &    Coef n1               & Coef n2           &  Coef n3            & Coef n0             & $\sigma$  \\
\hline
        QA    &   (Ks$-$A)o    &  $-$0.2611$\pm$0.0501   & $-$0.0003$\pm$0.0067 & $-$0.0001$\pm$0.0003   & 0.3172$\pm$0.1107    &  0.20\\
        QC    &   (Ks$-$C)o    &  $-$0.2057$\pm$0.0817   &    0.0047$\pm$0.0071 &    0.0000$\pm$0.0002   & 0.3386$\pm$0.2996    &  0.20\\
        QD    &   (Ks$-$D)o    &  $-$0.2694$\pm$0.0318   &    0.0039$\pm$0.0015 &        $..$            & 0.2591$\pm$0.1696    &  0.22\\
        QE    &   (Ks$-$E)o    &  $-$0.2327$\pm$0.0228   &    0.0020$\pm$0.0008 &        $..$            & 0.2781$\pm$0.1660    &  0.21\\

\hline
       QW3    &  (Ks$-$W3)o    & $-$0.1942$\pm$0.0271    & 0.0052$\pm$0.0021 &  0.0001$\pm$0.0000  & 0.3358$\pm$0.1038 & 0.20\\
       QW4    &  (Ks$-$W4)o    & $-$0.1006$\pm$0.0491    & 0.0126$\pm$0.0032 &  0.0003$\pm$0.0000  & 0.8190$\pm$0.2360 & 0.21\\
  
\hline
       Q58    &(Ks$$-$[5.8]$)o  & $-$0.3599$\pm$0.0414     &   0.0023$\pm$0.0128 &0.0000$\pm$0.0010  & 0.4767$\pm$0.0389 & 0.24\\
       Q80    &(Ks$$-$[8.0]$)o  & $-$0.3060$\pm$0.0416     &   0.0052$\pm$0.0093 &0.0004$\pm$0.0006  & 0.3977$\pm$0.0538 & 0.23\\
\hline
\end{tabular}
\begin{list}{}
  \item  {\bf Note:} Y-Fit = Coef~n1 $\times$ X-axis  + Coef~n2 $\times$ (X-axis)$^2$ + Coef~n3 $\times$ (X-axis)$^3$ +Coef~n0. \\
  The colours  used for the fit are those dereddened with the \Akint\ from surrounding stars
  by \citet{messineo05} and \citet{messineo18} rescaled to a power law with an index of $-2.1$
  and Gordon's extinction curve.
\end{list}
\end{table*}

\subsection{ \Akint\ and (\Ks$-[\lambda]_o$)  colours derived with \Ql\ and the 
\Ql-intrinsic-colours curves.}

The \Ql-intrinsic-colours curves derived in the above section
can be used to obtain estimates   of the 
(\Ks$-[\lambda])_o$  colours for  stars  with unknown parameters.

The following formula can be used to estimate the interstellar 
\Akint\ as a function of the observed and dereddened colours 
once the intrinsic colours have been identified with \Ql:

\begin{equation}
 A_{\it K_{\rm s}}{\rm (int)}= \frac{(Ks-[\lambda])-(Ks-[\lambda])_o}{1-\frac{A_{\lambda}}{A_{ K_{\rm s}}}}
.\end{equation}

The term $(Ks-[\lambda])_o$ results from 
Eq. \ref{equ1} and the coefficients from 
Tables \ref{table.qxfits} or
\ref{table.qxfits.gordon} by entering the  \Ql.
The term $(Ks-[\lambda])$ is the observed colour.
The  extinction ratios are listed in  Table  \ref{table:extinction}.

\begin{figure*}
\begin{center}
\resizebox{0.45\hsize}{!}{\includegraphics[angle=0]{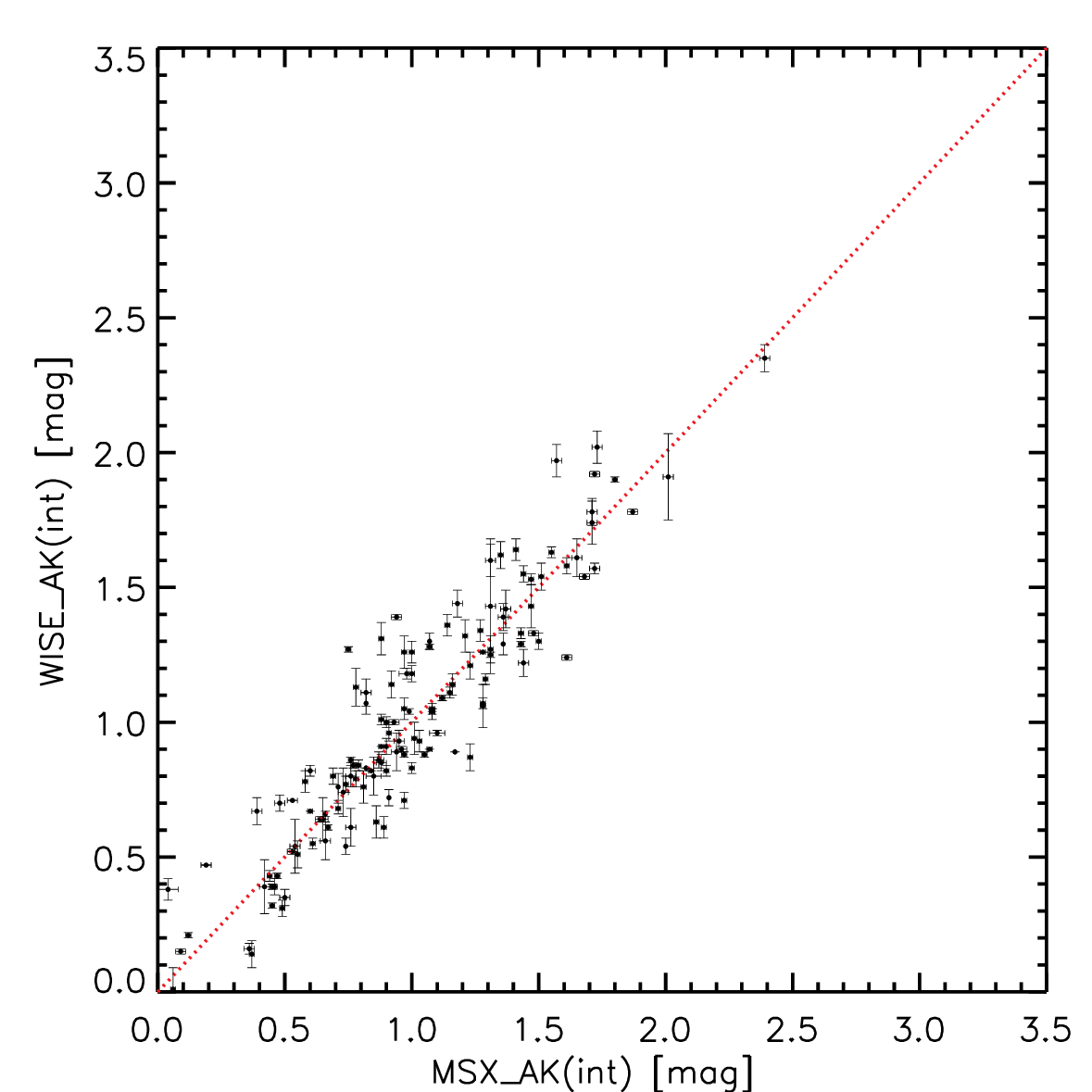}}
\resizebox{0.45\hsize}{!}{\includegraphics[angle=0]{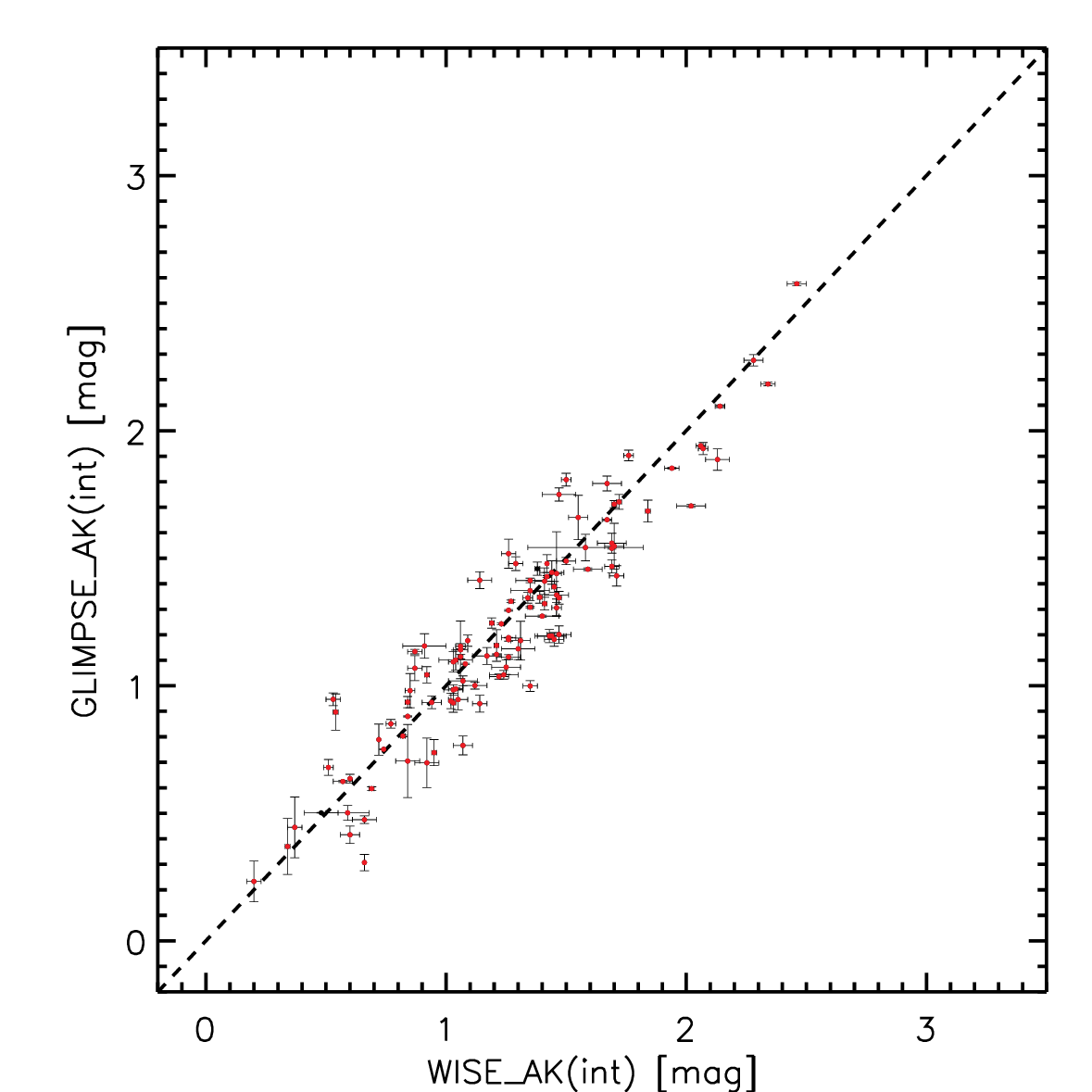}}
\end{center}
\caption{\label{akwisemsxgl} 
Comparison of \Akint\ derived from different datasets.
{\it Left:} \Akint\ values obtained from WISE data vs. 
those values from MSX data.
{\it Right:} \Akint\ values obtained from GLIMPSE data vs. 
those values from WISE data.
}
\end{figure*}

The  revised  curve 3 of \citet{messineo05} was used first.
The four independent \Akint\ obtained with the MSX magnitudes
were averaged; $\sigma=$ 0.05 mag.
We repeated the calculation with the WISE dataset and
the GLIMPSE dataset. The \Akint\ values from the  W3 and W4 WISE 
bands give  $\sigma$=0.03 mag, 
while the \Akint\ values from  GLIMPSE [5.8] and [8.0] data 
yield $\sigma$=0.03 mag.
There are 135 stars with good photometry and estimates of \Akint\ 
with MSX and WISE datasets; the average difference is 0.04 mag
and the $\sigma$ is 0.20 mag, as shown in Fig. \ref{akwisemsxgl}.
The 111 \Akint\  estimates of good photometric quality  
in common between  
the GLIMPSE and WISE  estimates yield  $\sigma$=0.15 mag.

Identical \Akint\ values are measured (within errors) after 
repeating the full computation using Gordon's effective 
ratios of Table \ref{table:extinction}.
The same \Aks\ values are obtained as long as the 
dereddened colours and the $Q_\lambda$
are performed consistently with the same interstellar
curve.
We measured \Akint\ by analyzing the correlations between \Ql\
and the intrinsic colours.
The \Akint\ values of individual stars are obtained independently of the
stellar environment, anywhere in the Galactic Disk.

\begin{figure*}
\begin{center}
\resizebox{0.33\hsize}{!}{\includegraphics[angle=0]{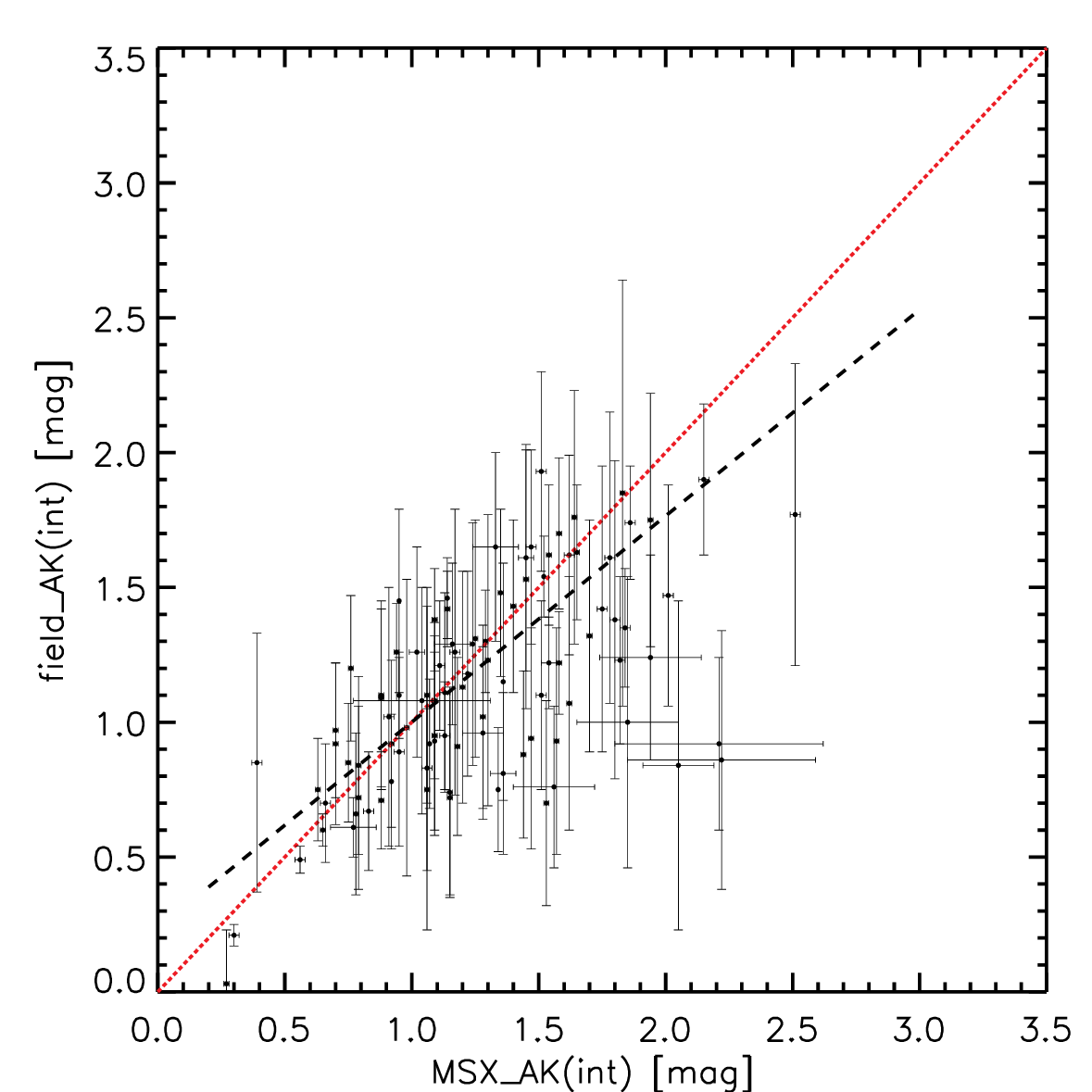}}
\resizebox{0.33\hsize}{!}{\includegraphics[angle=0]{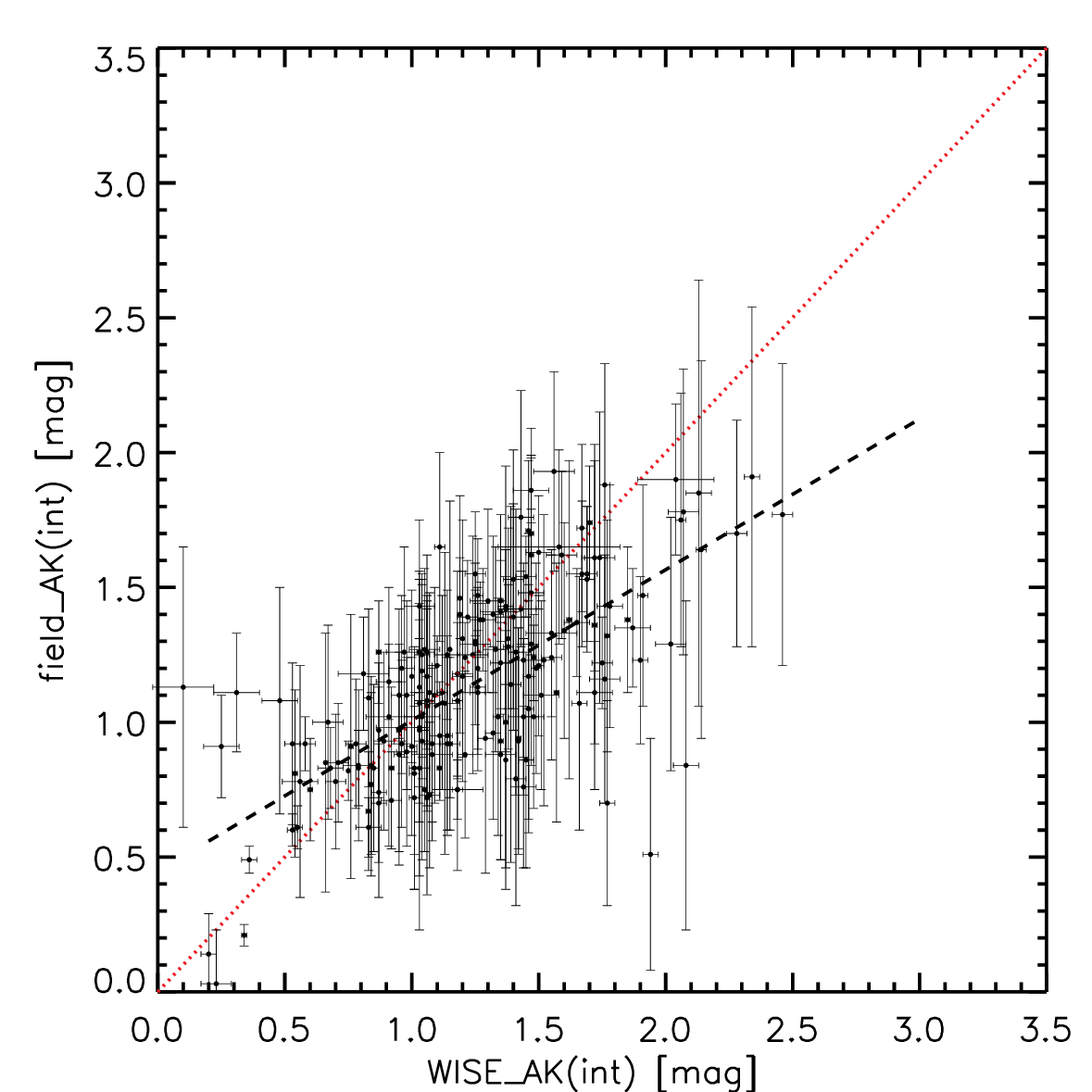}}
\resizebox{0.33\hsize}{!}{\includegraphics[angle=0]{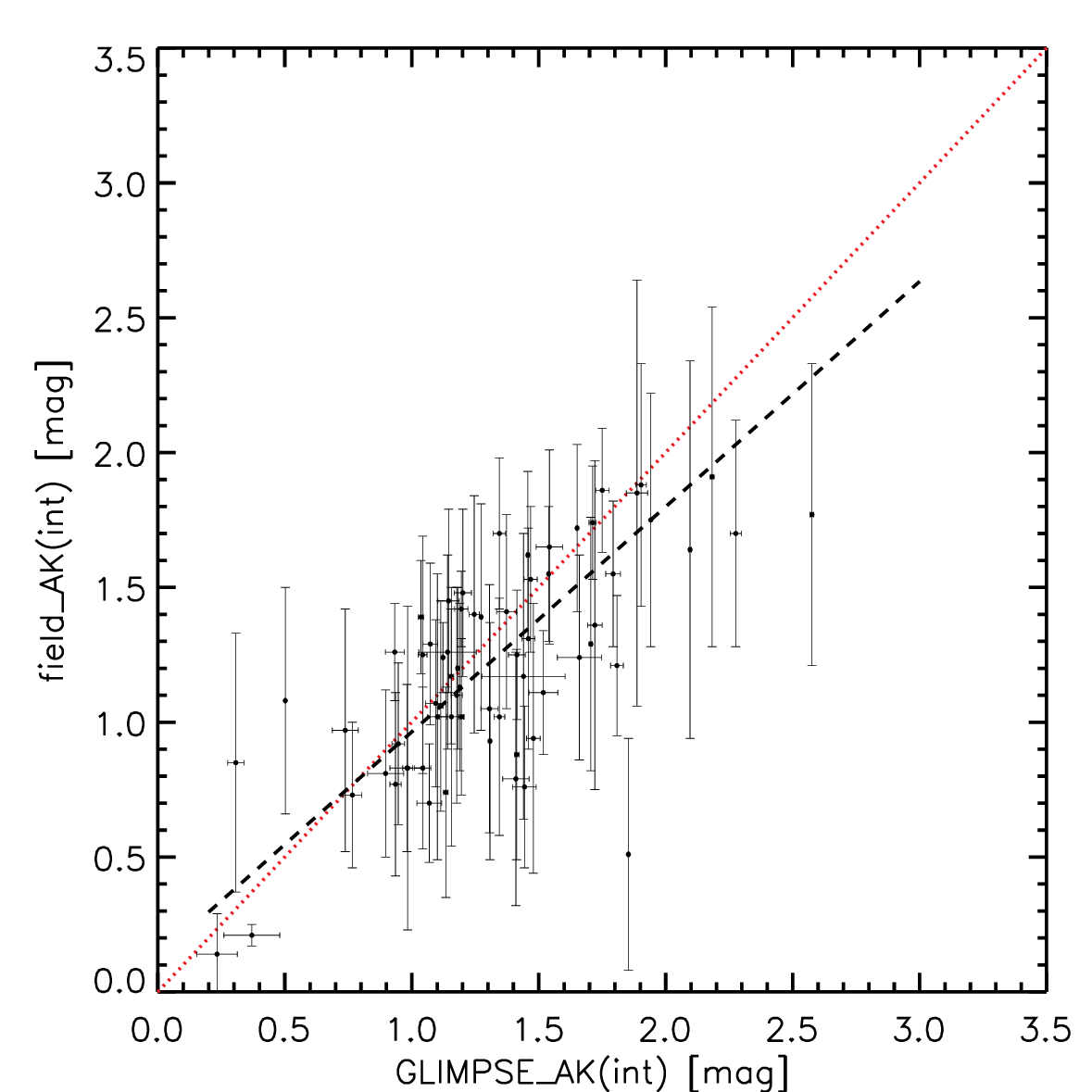}}
\end{center}
\caption{\label{ref.comp}  
Comparison of the three \Akint\ values estimated  from three MIR datasets
with the \Akint\ values from surrounding stars.
{\it Left panel:}  \Akint\ values from 
the MSX data are plotted vs. the median \Akint\ estimated with 
surrounding stars by \citet{messineo05}. {\it Middle panel:} 
 \Akint\ values from 
the WISE data are plotted vs. the median \Akint\ from \citet{messineo05}.
 {\it Right panel:} 
 \Akint\ values from 
the GLIMPSE data are plotted versus the median \Akint\ from \citet{messineo05}.
The dashed lines are fits to the data points (see text); 
the red dotted lines are the equity lines.
}
\end{figure*}

In Fig. \ref{ref.comp}, we compare the original estimate of \Akint\ 
(via the surrounding stars) and the one obtained with 
the fits and \Ql.
Using the MSX data, the average \Akint(fit)-\Akint(surrounding)=0.14 mag
with a $\sigma=0.36$ mag and a median of 0.07 mag;  using the WISE 
data the average \Akint(fit)-\Akint(surrounding)=0.04 mag
with a $\sigma=0.32$ mag and a median  of 0.00 mag;  
using the GLIMPSE data \Akint(fit)-\Akint(surrounding)=0.05 mag
with a $\sigma=0.33$ mag and a median  of 0.03 mag.
The data points displayed in Fig. \ref{ref.comp} were fitted with a Huber regression;
for the GLIMPSE and the MSX data points, at \Aks=2 mag the fits indicate
values 0.2 mag larger than the equity line, suggesting that
the new estimated \Aks\ magnitudes may be  0.2 mag
larger than those estimated by \citet{messineo05}.
This systematic error could be corrected a posteriori.
However, as shown in the discussion section, for  Miras at the Galactic center
the new method yields
\Aks\ values that are 0.1 mag smaller than comparison values from the literature.
The regression line for the WISE dataset shows a larger offset from 
the equity line of 0.4 mag at \Aks = 2 mag.
This  method allows for an accuracy of $\approx 0.35$ mag,
which corresponds to the scatter
of the data points  in the diagram $J-$\Ks\ versus \Ks$-D$  
of Fig.\ \ref{fig.bcD} (the NIR and MIR measurements
were not acquired simultaneously).

\subsection{  Estimates of  \Akenv\ as \Aktot-\Akint }
\label{Akenvelope}

The $H-$\Ks\ colours of  naked giants were  calculated  
by synthetic photometry of 
theoretical spectra.
For this purpose, the publicly available  
NextGen spectra of \citet[][]{allard11} 
with \Teff=3500, 3200, 2600 K, log$_{10}$(g)=1,
and solar metallicity were used\footnote{The 
spectra are distributed by the Virtual
Observatory SED analyzer \citep{bayo08}.
The bt-nextgen\_agss2009 (gas only) were retrieved.}.
The synthetic spectra were generated with the PHOENIX atmospheric code
 considering spherical symmetry.
The synthetic spectra  reproduce  the TiO and VO molecules, 
which dominate  the optical spectrum, 
and water vapour and CO molecules
in the infrared spectrum. The molecular list was refined to 
reproduce the atmospheric lines of brown dwarfs 
\citep{allard11,allard14}.
The infrared colours  are obtained by convolving
the spectrum with the filter
and calculating the mean flux density.
The adopted zero points come from \citet[][2MASS]{cohen03},
 \citet[][MSX]{egan03}, \citet[][WISE]{jarrett11},
and \citet[][GLIMPSE]{fazio04}.
The synthetic infrared colours are listed
in Table \ref{naked}.

The average differences between the \Aks(tot) calculated with 
the synthetic ($H-$\Ks) colours
and those obtained with the colours of \citet{koornneef83}
are +0.05 mag with $\sigma$=0.00 mag (M4) and +0.14 mag
with $\sigma$=0.00 mag (M6).
While the average differences between the \Aks(tot) 
calculated with 
the synthetic ($H-$\Ks) colours for a 3500 K (M4) and a 3200 K (M5-6) star
are +0.015 mag with $\sigma$=0.006 mag.
The   \Aks(int) values
were re-determined using 
the fiducial color sequences 
Q$_\lambda \propto {\rm (K_S-\lambda)_o}$,
as described in Sec. \ref{Ql}.

\begin{figure}
\begin{center}
\resizebox{0.8\hsize}{!}{\includegraphics[angle=0]{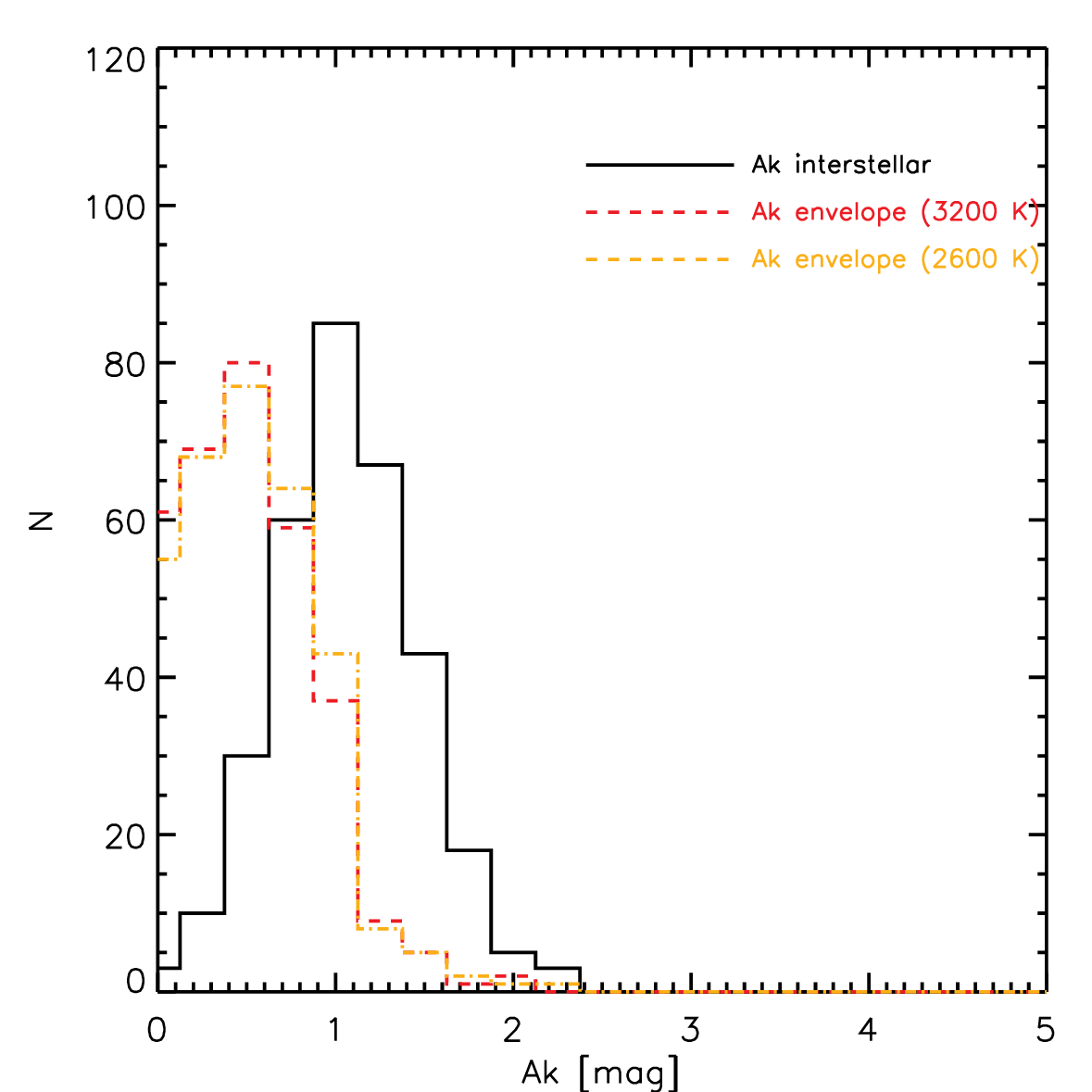}}
\end{center}
\caption{\label{histo_Ak} Histogram of the obtained interstellar \Ak\ 
values (black line). 
 The envelope \Ak\ values (\Akenv=\Aktot$-$\Akint)
are superimposed (in red using the colours of a 
3200 K naked star and in orange with the colours of a 
2500 K naked star).}
\end{figure}

Using the derived  \Aktot\ and \Akint, we could estimate
the envelope extinction \Akenv\ for  324 targets.
\Akenv=\Aktot$-$\Akint\ ranges from 0 to 2.5 mag,
as shown in Fig. \ref{histo_Ak}.
The envelope's optical depth at 0.55 \um, $\tau_{0.55}$, ranges from 0 to 32
when estimating it as: $\approx $\Akenv$/ 0.077$\footnote{  \Av/\Aks = 12.99 is appropriated 
for an interstellar infrared power law of 
index  = $-2.1$ and R$_{\rm V}$ = 3.1 \citep{messineo05}. 
This conversion factor is close to the 12.82 obtained by \citet{wang19}
for Galactic extinction. As shown in the next section, for a circumstellar
extinction law, the actual A$_{0.55}$/A$_{2.2}$ ratio depends on the adopted grain sizes. 
It is 15.7 for a maximum size of 0.25 \um\ and 3.7 for a maximum size of 0.75 \um.}.

\begin{table}
\caption{\label{naked} Naked star colours}
\begin{tabular}{llrrr}
\hline
\hline
{\it T$_{\rm eff}$}& [K] &  3500     & 3200& 2600 \\
Spectral type$^a$ &      & M4        & M5-6&  M9$^c$  \\
$J-K_{\rm s}^b$   & [mag]&  1.198    &1.245&0.967\\
$H-K_{\rm s}^b$   & [mag]&  0.231    &0.244&0.222\\
$K_{\rm s}-K_{\rm s}^b$& [mag]& 0.000&0.000&0.000\\
$K_{\rm s}-[3.6]$ & [mag]& 0.115     &0.104&0.320\\
$K_{\rm s}-[4.5]$ & [mag]& $-0.004$  &0.038&0.449\\
$K_{\rm s}-[5.8]$ & [mag]& 0.032     &0.051&0.473\\
$K_{\rm s}-[8.0]$ & [mag]& 0.194     &0.161&0.596\\
$K_{\rm s}-A$     & [mag]& 0.052     &0.041&0.504\\
$K_{\rm s}-C$     & [mag]& 0.208     &0.228&0.796\\
$K_{\rm s}-D$     & [mag]& 0.216     &0.242&0.822\\
$K_{\rm s}-E$     & [mag]& 0.225     &0.275&0.888\\
$K_{\rm s}-W1$    & [mag]& 0.081     &0.009&0.151\\
$K_{\rm s}-W2$    & [mag]& $-0.031$  &0.006&0.425\\
$K_{\rm s}-W3$    & [mag]& 0.207     &0.219&0.759\\
$K_{\rm s}-W4$    & [mag]& 0.218     &0.271&0.887\\
\hline
\end{tabular}
\begin{list}{}
\item The naked colours are based on  models of stars with log(g)=1 
and solar metallicity by \citet{allard11}.
\item ($^a$) Spectral types are based  on the temperature scale of \citet{vanbelle21}.\\
\item ($^b$) For  M4, M5, and M6 stars, intrinsic  $J-K_{\rm s}$ colours 
of 1.054, 1.164 and 1.208 mag, respectively, were measured by \citet{glass02}.
While, \citet{koornneef83} report $J-K$ = 1.16 mag for an M4 and 
1.30 for an M6, and $H-K$ = 0.27 mag and 0.35  mag.
\item ($^c$)\citet{perrin98}
\end{list} 
\end{table}

\subsection{ Empirical circumstellar excess colors.}

\begin{figure*}
\begin{center}
\resizebox{0.24\hsize}{!}{\includegraphics[angle=0]{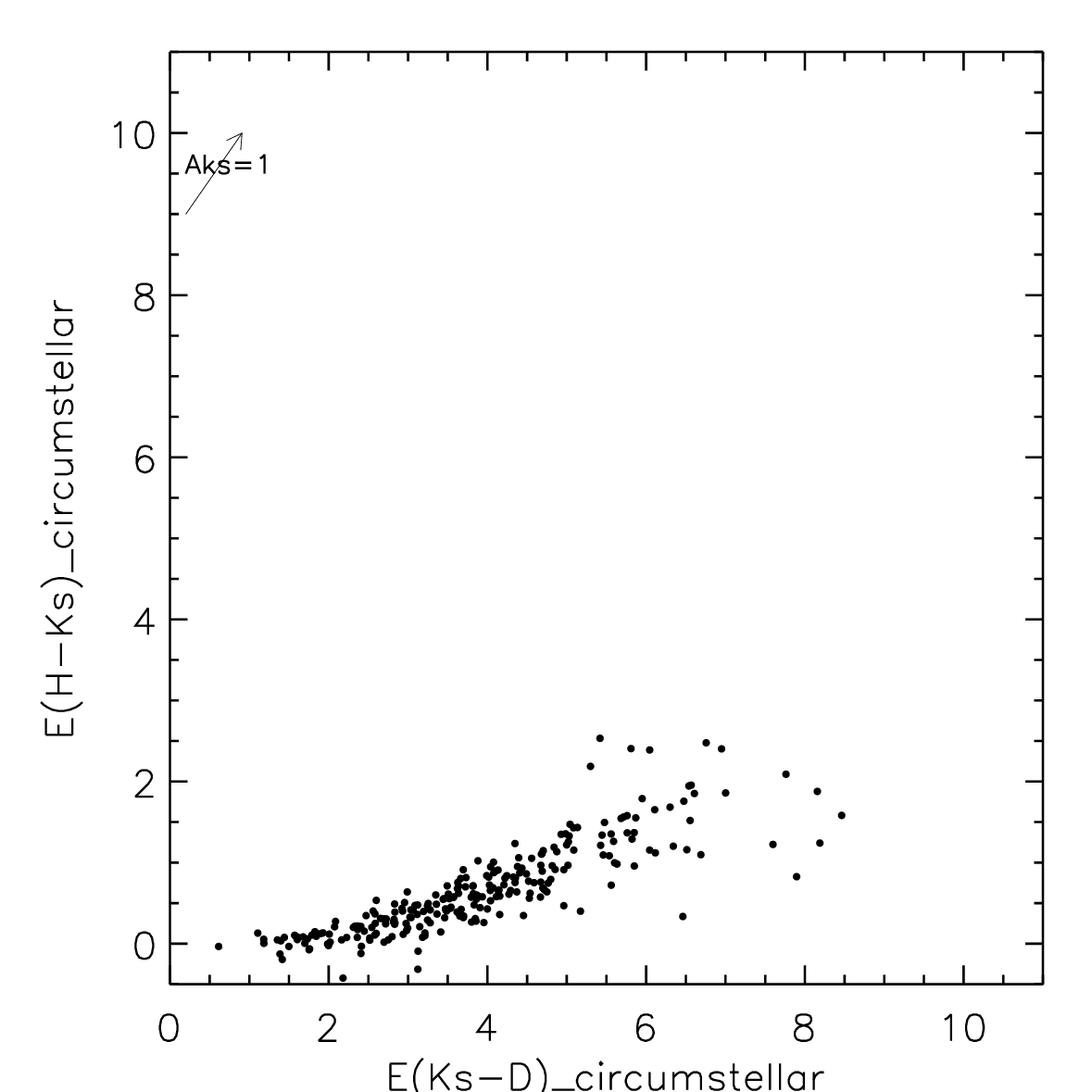}}
\resizebox{0.24\hsize}{!}{\includegraphics[angle=0]{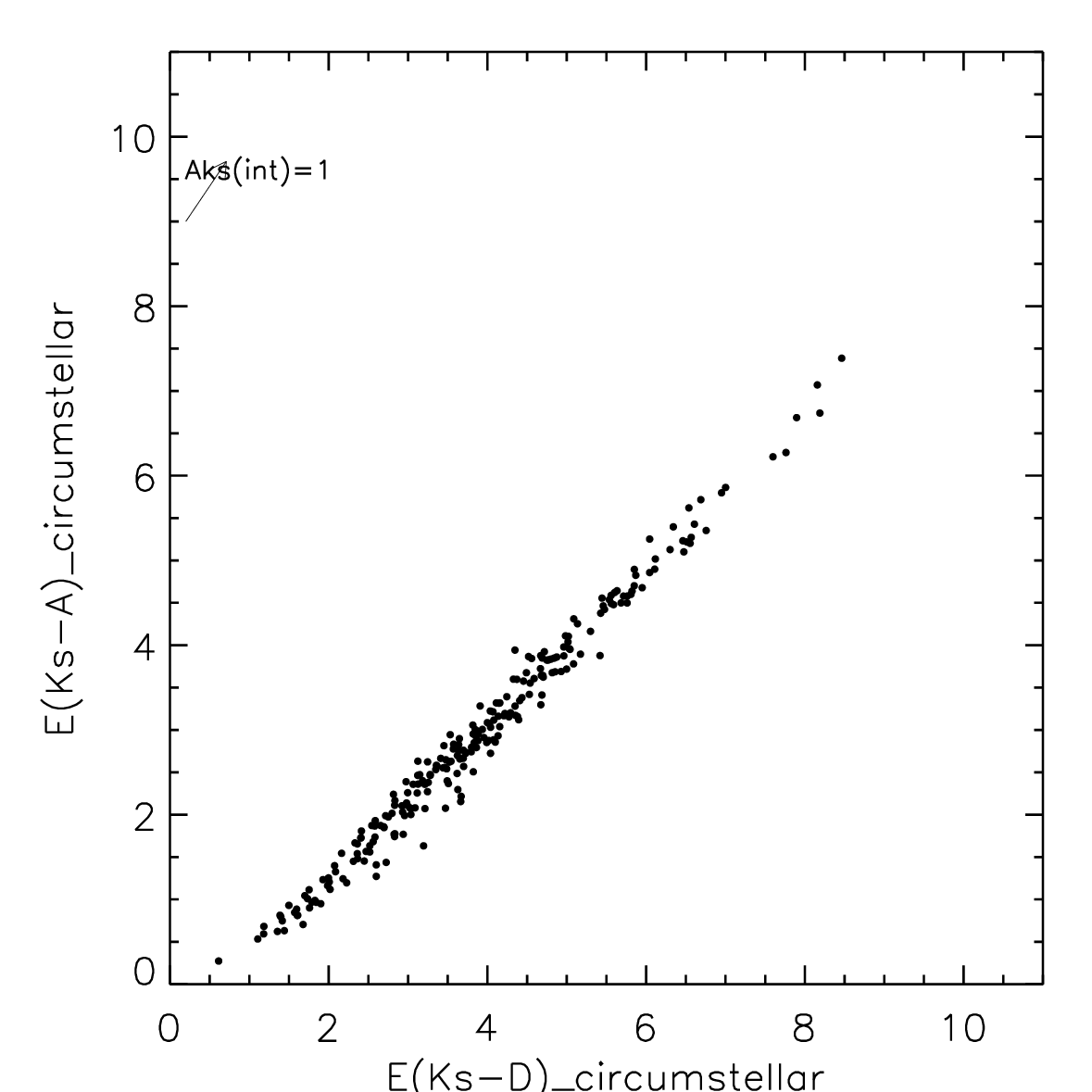}}
\resizebox{0.24\hsize}{!}{\includegraphics[angle=0]{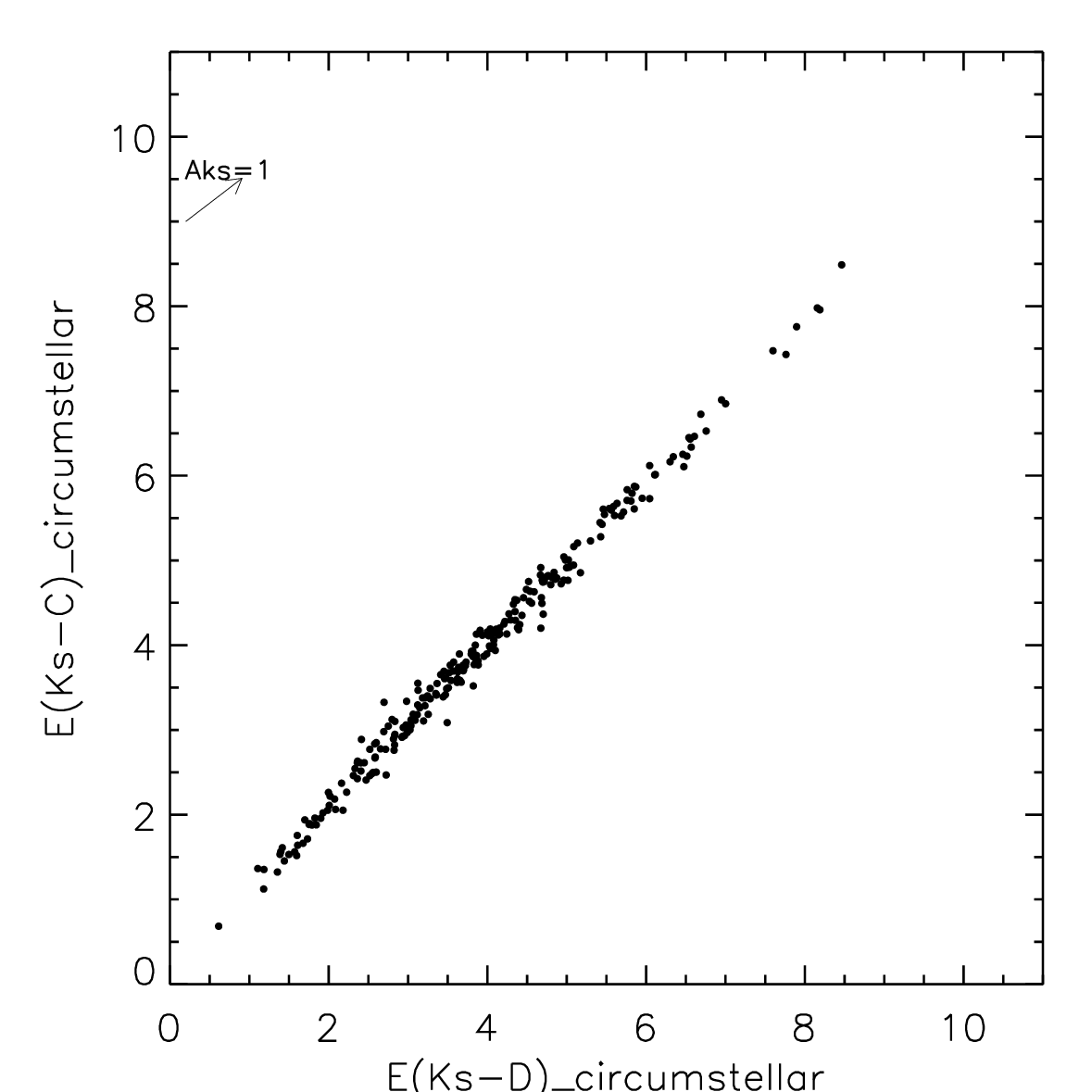}}
\resizebox{0.24\hsize}{!}{\includegraphics[angle=0]{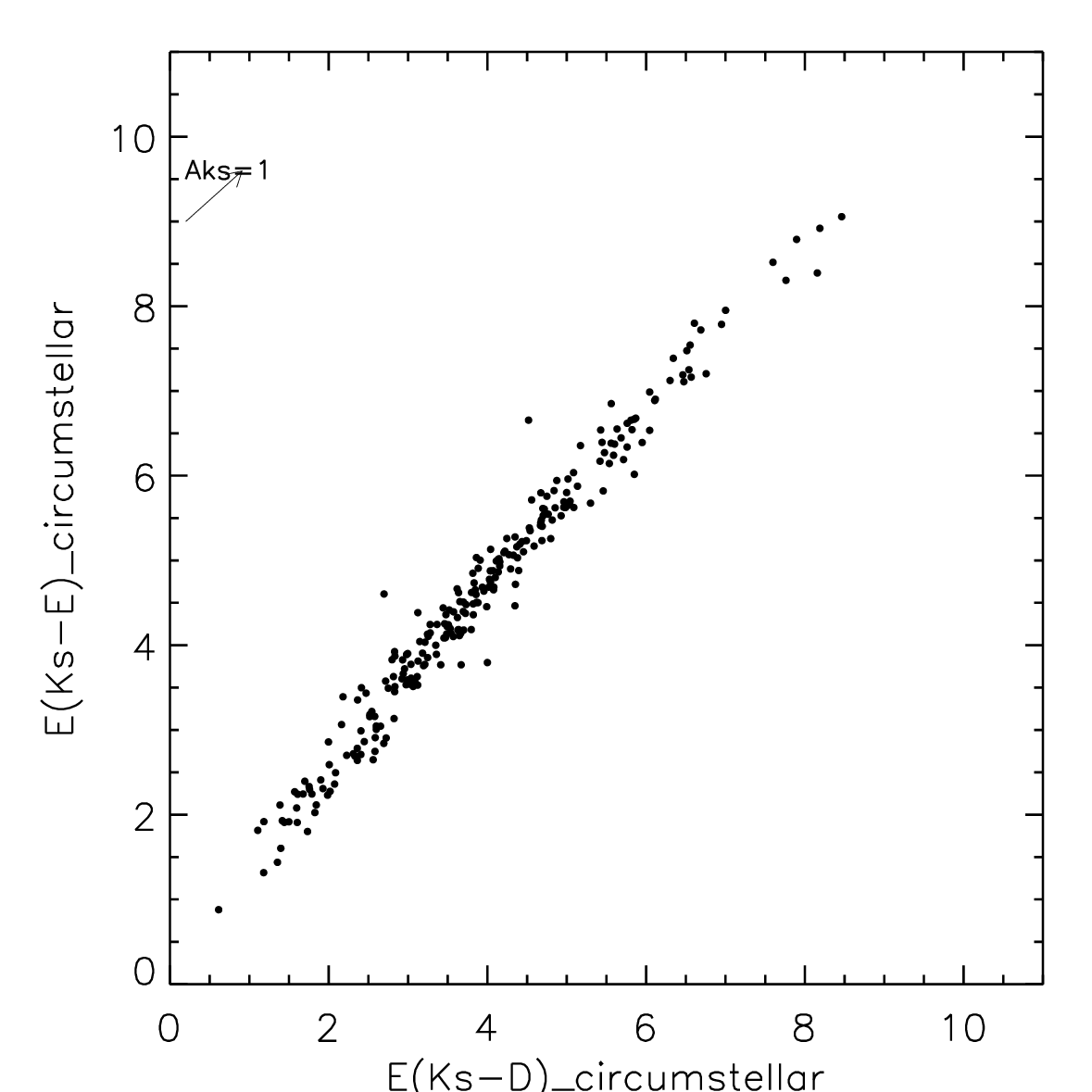}}
\end{center}
\caption{\label{circ.excessmsx}  Circumstellar color excess of the sample stars
with 2MASS and MSX colours (based on the naked colours of a 3200 K star).}
\end{figure*}

\begin{figure*}
\begin{center}
\resizebox{0.24\hsize}{!}{\includegraphics[angle=0]{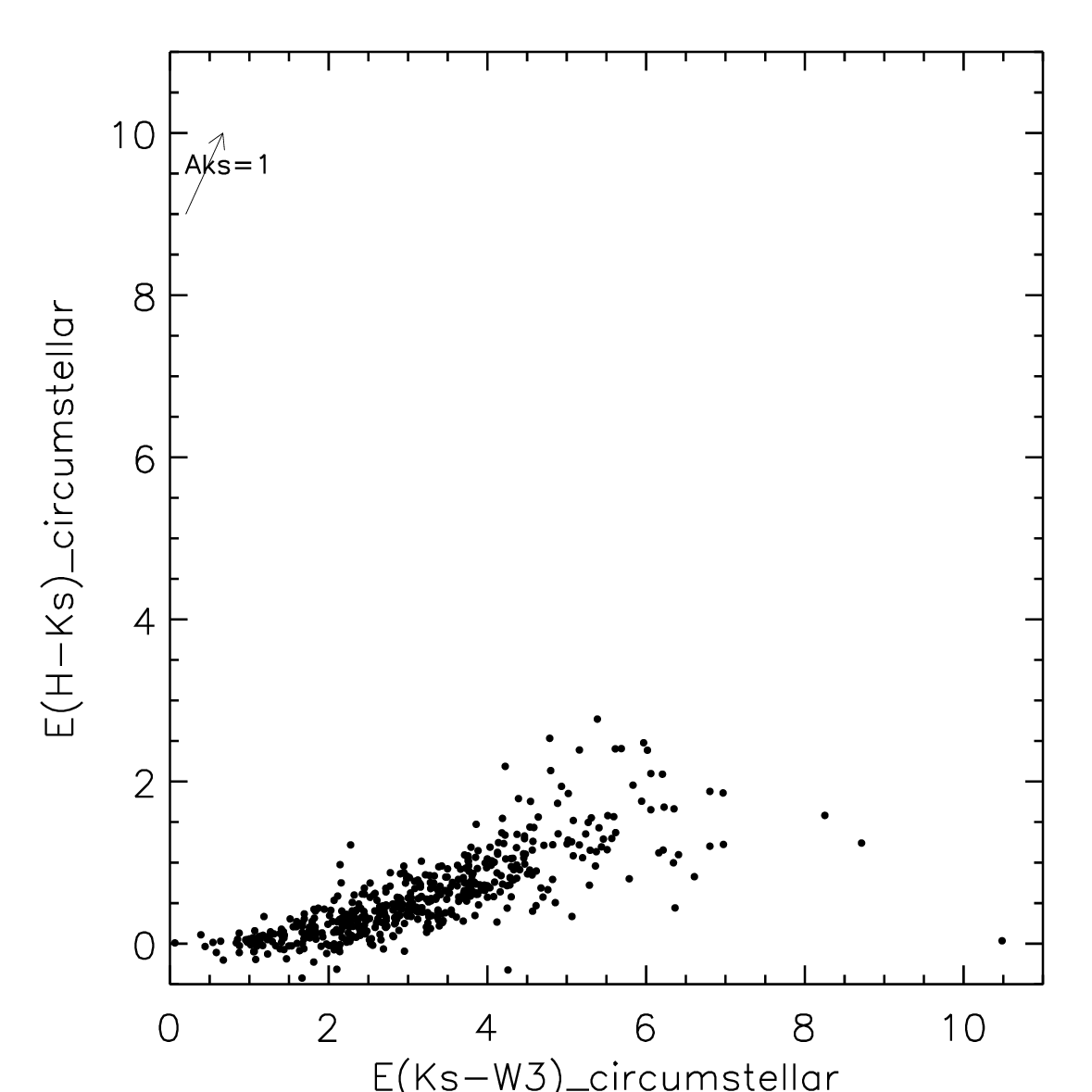}}
\resizebox{0.24\hsize}{!}{\includegraphics[angle=0]{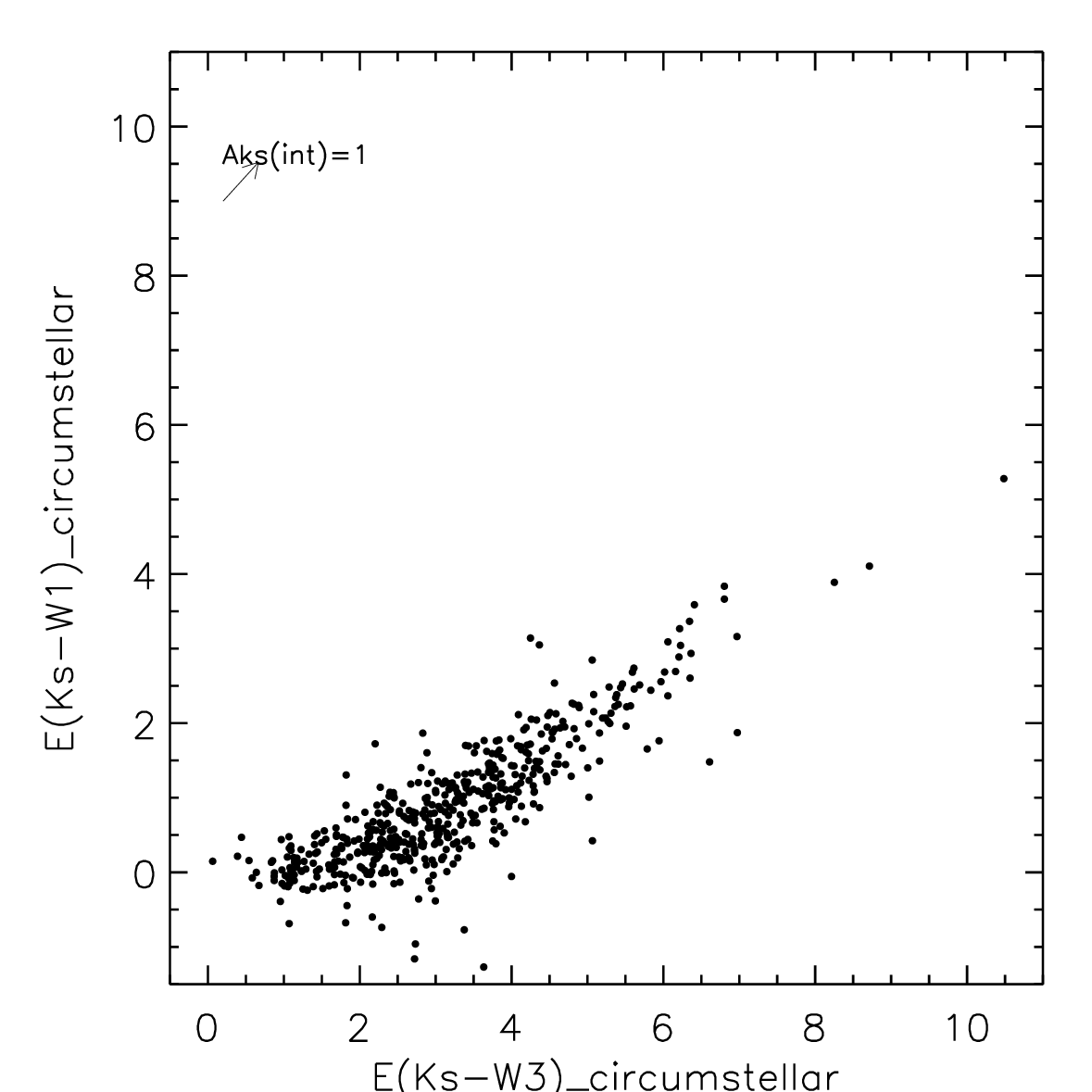}}
\resizebox{0.24\hsize}{!}{\includegraphics[angle=0]{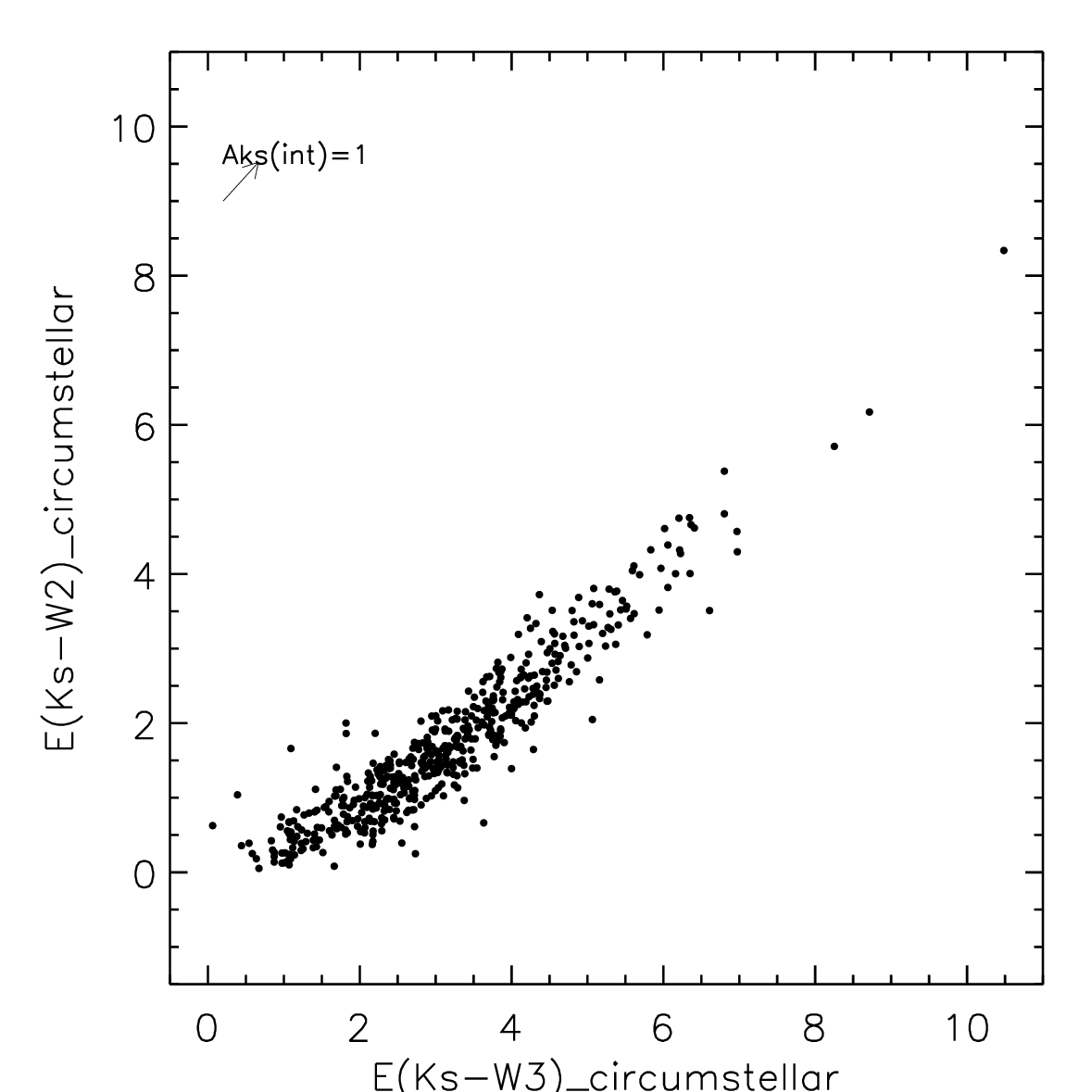}}
\resizebox{0.24\hsize}{!}{\includegraphics[angle=0]{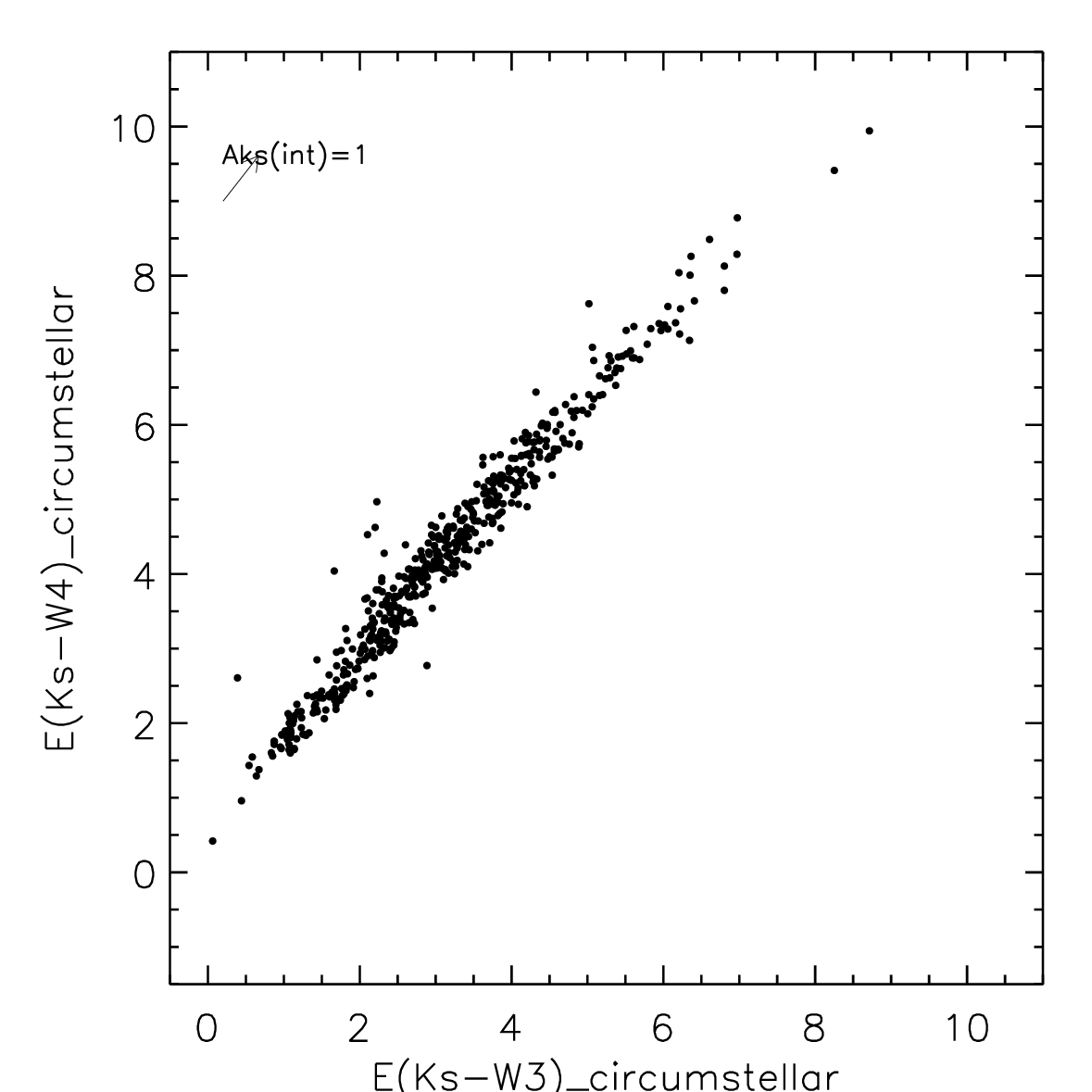}}
\end{center}
\caption{\label{circ.excessWISE}   Circumstellar color excess of the sample stars
with 2MASS and WISE colours (based on the naked colours of a 3200 K star).}
\end{figure*}

\begin{figure*}
\begin{center}
\resizebox{0.24\hsize}{!}{\includegraphics[angle=0]{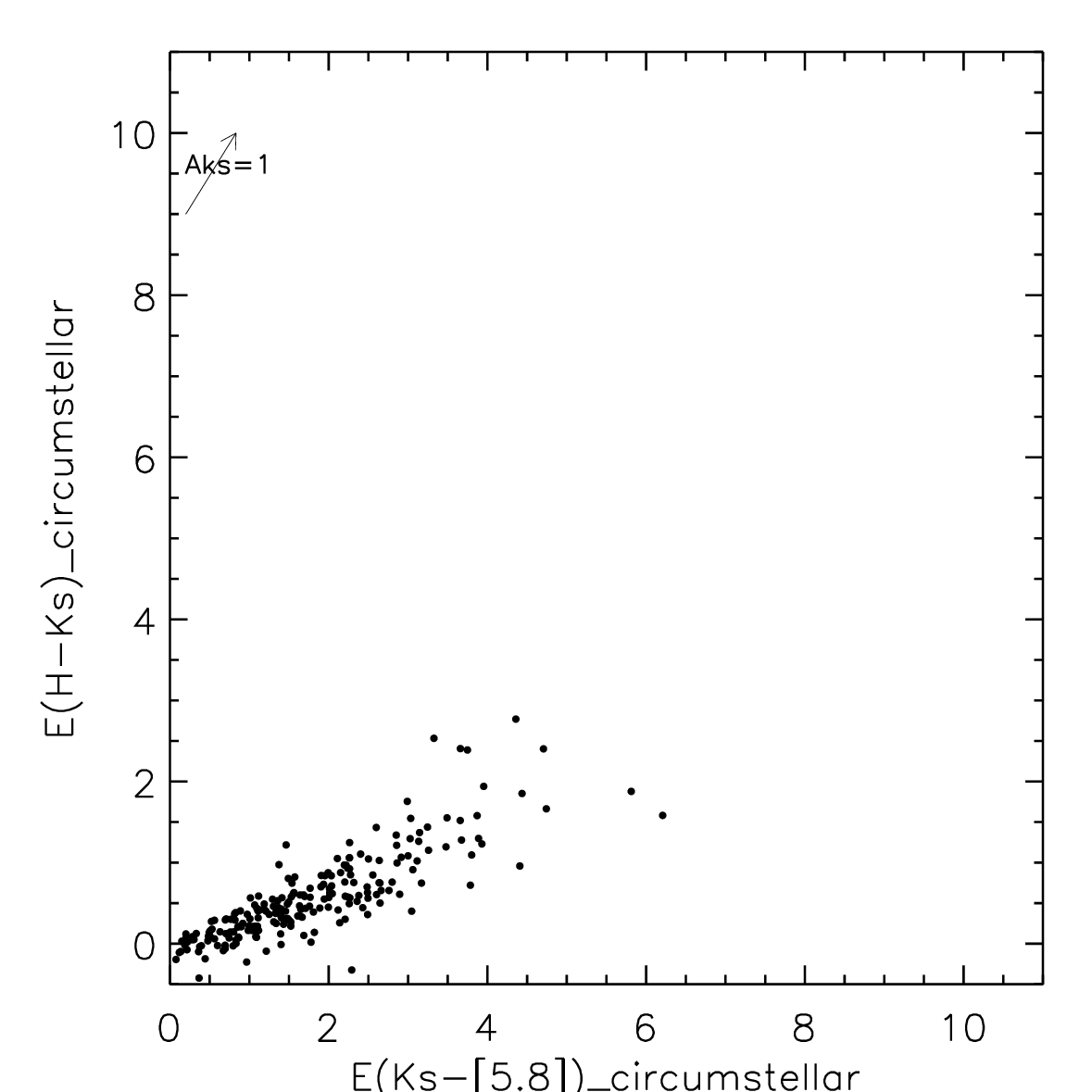}}
\resizebox{0.24\hsize}{!}{\includegraphics[angle=0]{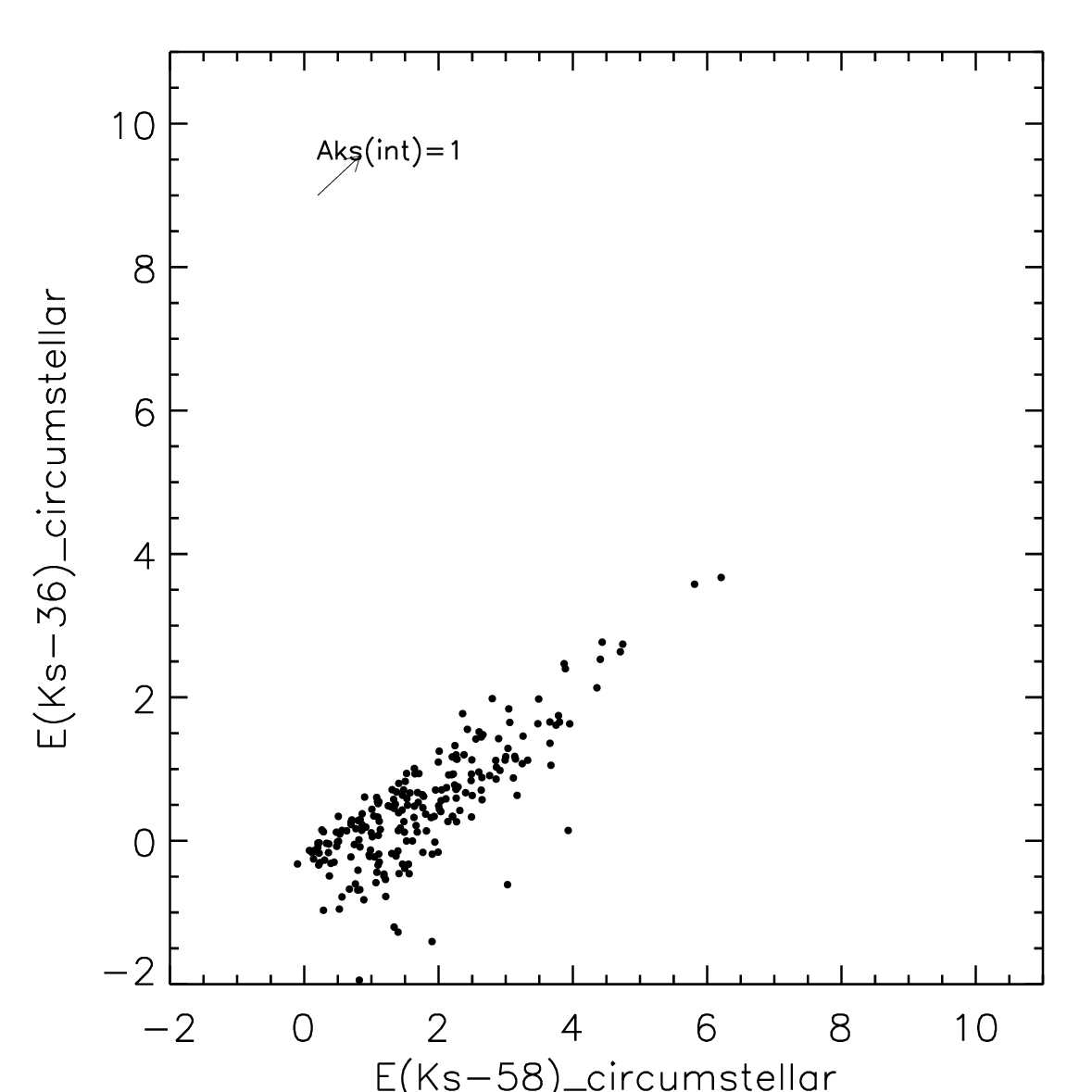}}
\resizebox{0.24\hsize}{!}{\includegraphics[angle=0]{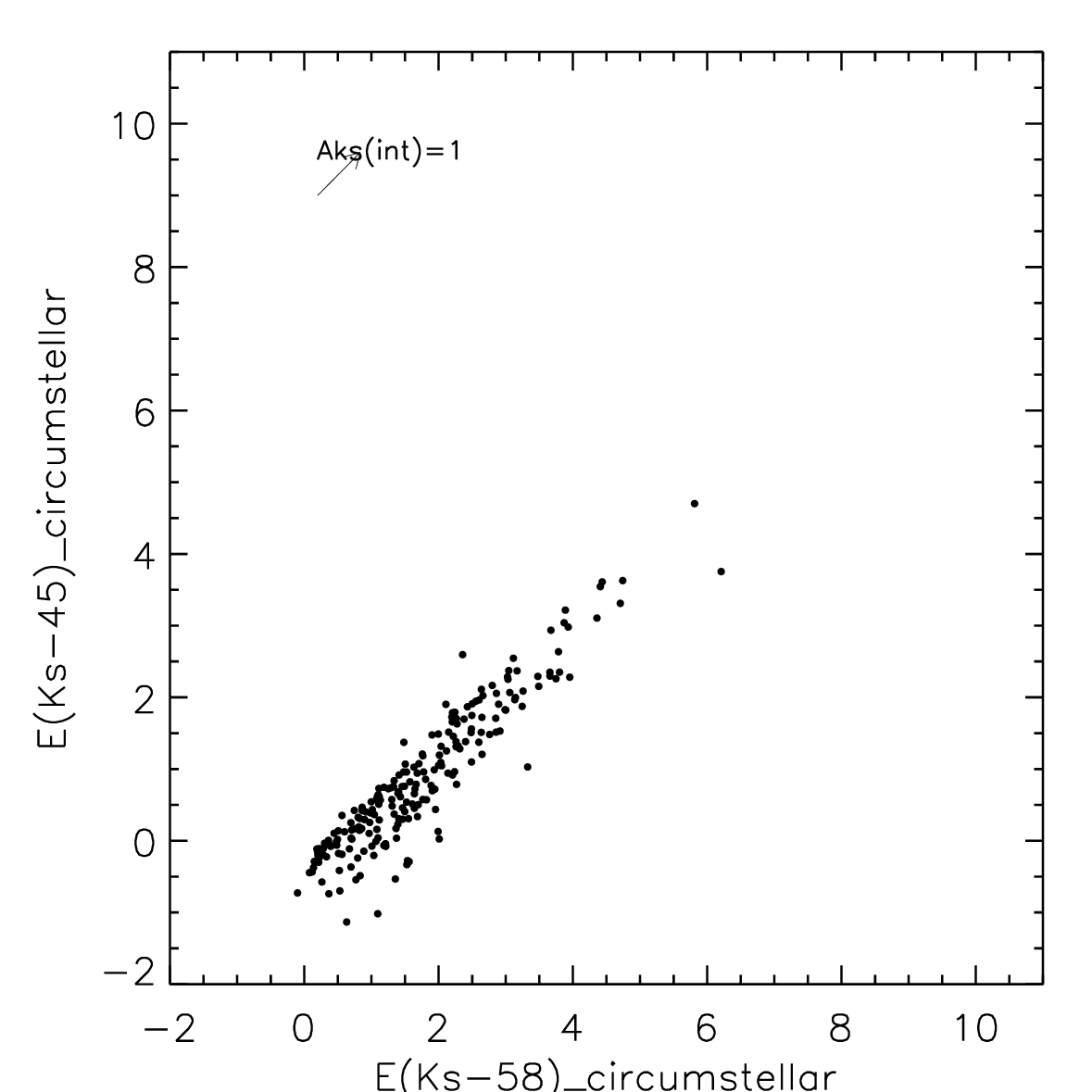}}
\resizebox{0.24\hsize}{!}{\includegraphics[angle=0]{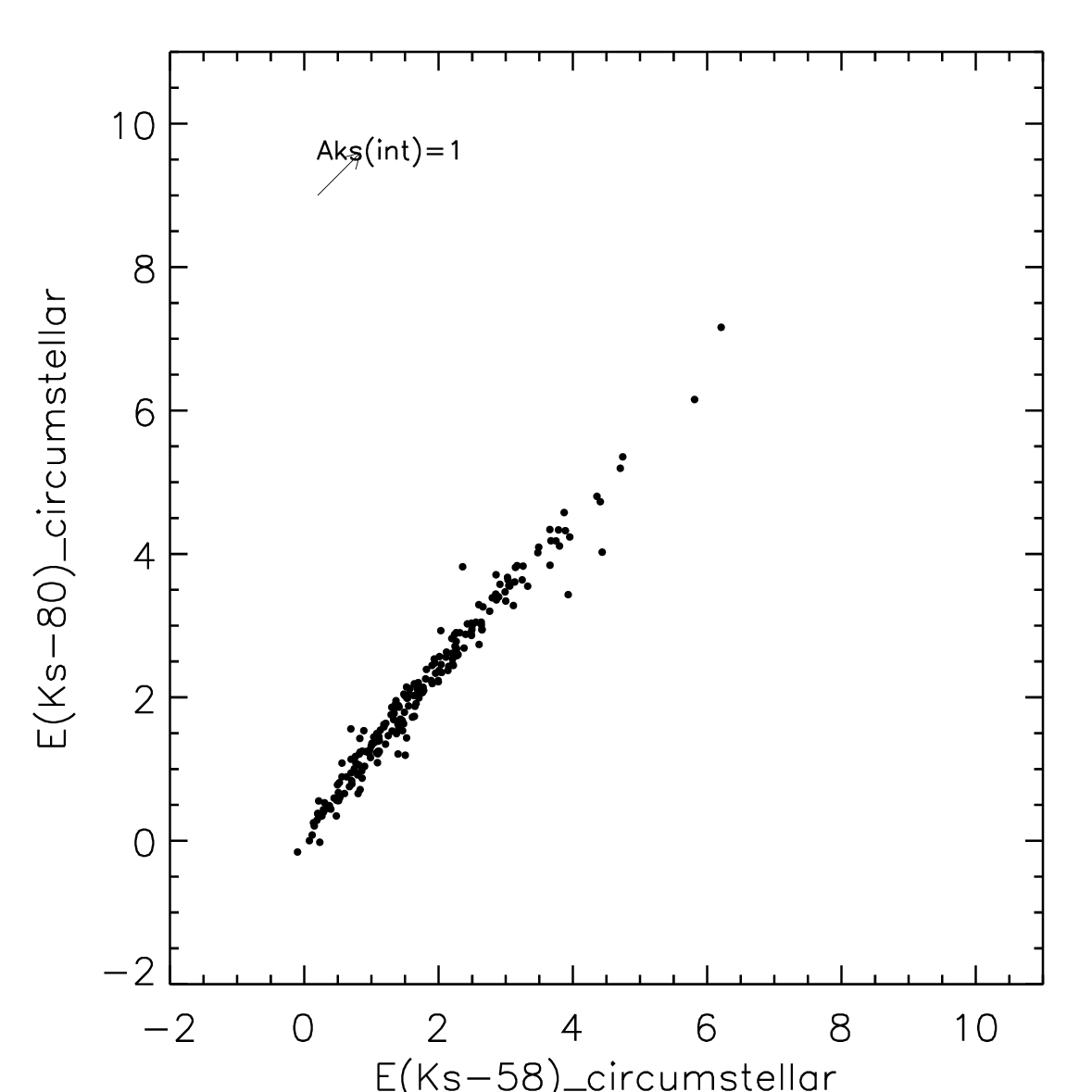}}
\end{center}
\caption{\label{circ.excessGLIMPSE} The circumstellar colour excess of the sample stars
with 2MASS and GLIMPSE colours (based on the naked colours of a 3200 K star).}
\end{figure*}

\begin{table}
\caption{\label{table_exratio}  Observed circumstellar excess ratios, 
based on the naked colours of a 3200 K star.}
\begin{tabular}{llllll}
\hline
\hline
Excess ratios &                    mean\_1  & sigma\_1 & mean\_2  & sigma\_2 \\
              &                    (Curve3) & (Curve3) & (Gordon) & (Gordon) \\
\hline
$\frac{\bf E(Ks-A)}{\bf E(Ks-D)}$        &{\bf  0.73} &{\bf 0.10}         & {\bf 0.70} & {\bf 0.11}\\
$\frac{\bf E(Ks-C)}{\bf E(Ks-D)}$        &{\bf  1.02} &{\bf 0.05}         & {\bf 0.97} & {\bf 0.04}\\
$\frac{\bf E(Ks-E)}{\bf E(Ks-D)}$        &{\bf  1.20} &{\bf 0.10}         & {\bf 1.18} & {\bf 0.10}\\
$\frac{E(Ks-W1)}{E(Ks-W3)}$      & 0.24 & 0.22         & 0.31 & 0.23\\
$\frac{E(Ks-W2)}{E(Ks-W3)}$      & 0.54 & 0.46         & 0.58 & 0.54\\
$\frac{E(Ks-W4)}{E(Ks-W3)}$      & 1.44 & 0.39         & 1.50 & 0.47\\
$\frac{E(Ks-[3.6])}{E(Ks-[5.8])}$& 0.08 & 0.67         & 0.23 & 0.58\\
$\frac{E(Ks-[4.5])}{E(Ks-[5.8])}$& 0.28 & 0.89         & 0.31 & 0.91\\
$\frac{\bf E(Ks-[8.0])}{\bf E(Ks-[5.8])}$&{\bf  1.24} &{\bf 0.25}         & {\bf 1.20} & {\bf 0.30}\\
\hline
\end{tabular} 
\begin{list}{}
\item  The mean\_1 and sigma\_1 values are obtained after having dereddened
the observed colours with the \Aks\ values 
and the  revised  Curve 3 of \citet{messineo05}.
\item  The mean\_2 and sigma\_2 values are obtained after having dereddened
the observed colours with the \Aks\ values 
of \citet{messineo05}  (rescaled to a power law of index $-2.1$)
and the mid-interstellar extinction Curve  by \citet{gordon21}.
\end{list}
\end{table}

\begin{table}
\caption{\label{table_exratio2600}  Observed circumstellar excess ratios, based on the naked colors of a 2600 K star.}
\begin{tabular}{llllll}
\hline
\hline
Excess ratios &                    mean\_1  & sigma\_1 &  mean\_2  &  sigma\_2\\
              &                    (Curve3) & (Curve3) &  (Gordon) & (Gordon)\\
\hline
$\frac{\bf E(Ks-A)}{\bf E(Ks-D)}$        &{\bf  0.68}  &{\bf 0.43}&{\bf 0.65} &{\bf 0.32} \\
$\frac{\bf E(Ks-C)}{\bf E(Ks-D)}$        &{\bf  1.04}  &{\bf 0.17}&{\bf 0.97} &{\bf 0.05} \\
$\frac{\bf E(Ks-E)}{\bf E(Ks-D)}$        &{\bf  1.26}  &{\bf 0.46}&{\bf 1.22} &{\bf 0.24} \\
$\frac{E(Ks-W1)}{E(Ks-W3)}$              &0.34         & 0.78     & 0.27 & 0.39 \\
$\frac{E(Ks-W2)}{E(Ks-W3)}$              &0.72         & 0.53     & 0.44 & 0.51 \\
$\frac{E(Ks-W4)}{E(Ks-W3)}$              &3.19         & 36.22    & 1.54 & 1.77\\
$\frac{E(Ks-[3.6])}{E(Ks-[5.8])}$        &0.05         & 2.97     & 0.45 & 4.69\\
$\frac{E(Ks-[4.5])}{E(Ks-[5.8])}$        &0.30         & 2.91     & 0.43 & 3.93\\
$\frac{\bf E(Ks-[8.0])}{\bf E(Ks-[5.8])}$&{\bf 1.30}   &{\bf 0.80}&{\bf 1.18} &{\bf 1.46}\\
\hline
\end{tabular} 
\begin{list}{}
\item  The mean\_1 and sigma\_1 values are obtained after having dereddened
the observed colors with the \Aks\ values 
and the  revised curve 3 of \citet{messineo05}.
\item  The mean\_2 and sigma\_2 values are obtained after having dereddened
the observed colours with the \Aks\ values of \citet{messineo05}
(rescaled to a power law of index $-2.1$)
and the mid-interstellar extinction curve  by \citet{gordon21}.
\end{list}
\end{table}

The naked colours listed in Table \ref{naked}
can be useful for determining the  colour excess ratios
and to set constraints on the circumstellar extinction curve.

A circumstellar colour excess is defined as
\begin{equation}
E(Ks-[\lambda])_{\rm env}=(Ks-[\lambda])_o-(Ks-[\lambda])_{\rm naked}
,\end{equation}
where $(Ks-[\lambda])_o$ is the interstellar-dereddened colour
and $(Ks-[\lambda])_{\rm naked}$ is the expected color for a
star of similar parameters (e.g., gravity, metallicity, and temperature),
but without an envelope.
For the $(Ks-[\lambda])_{\rm naked}$ values,
the synthetic colors are taken  from a stellar 
model with a temperature of 3200 K  
(Table \ref{naked}).

In Fig. \ref{circ.excessmsx},  the E(Ks-[D]) 
colour excess is plotted  against the  E(K-[A]),  
E(Ks-[C]), and E(Ks-[E]) color excess.
Figs. \ref{circ.excessWISE}
and \ref{circ.excessGLIMPSE} show similar 
diagrams for the WISE and GLIMPSE filters.
There is a tight correlation between the MSX excess colours, 
as well as between the E(Ks-[W4]) and E(Ks-[W3]) excess colors
and the  E(Ks-[8.0]) and E(Ks-[5.8]) excess colors.
On the contrary, a large scatter appears in the 
diagrams showing the E(Ks-[W1]), E(Ks-[W2]), E(Ks-[3.6]),
and E(Ks-[4.5]) excess colours in Figs.  \ref{circ.excessWISE}
and \ref{circ.excessGLIMPSE}.
Indeed, the stellar spectrum of cold stars at wavelenghts shorter than  
4 \um\ is dominated by strong 
molecular bands \citep{bedijn87}. 
Furthermore, the data  used here are from a
single-epoch, and the NIR\ and MIR data  were not taken simultaneously. 
The radial pulsation expands the atmosphere,
and creates an extra layer of molecules
(CO, H$_2$O, CO$_2$, and SiO) above the atmosphere. 
The region from 
2.4 to 4 \um\ is dominated by strong water vapour, 
which is the main source of opacity.
The water vapour absorption changes dramatically  with the
pulsation phase \citep{matsuura02}. 
This extended layer of molecules was
analyzed in detail in the work of  \citet{matsuura02}
and \citet{cami02}.
The larger scatter of data points shown in Figs. 
\ref{circ.excessWISE} and \ref{circ.excessGLIMPSE},
with the filters W1 and W2, and with the 
filters [3.6] and [4.5],
indicates the presence of strong absorption by water.

A tighter correlation of the  colour excess 
E(\Ks-D) with E(\Ks-A), E(\Ks-C), and E(\Ks-E)
confirms that the filters A, C, D, and E are 
dominated by dust emission, 
as well as W3, W4, [5.8], and [8.0].  
As already  reported by \citet{volk88} for the LRS spectra: 
``in the region from 8-23 \um\ the emergent
spectrum will directly reflect the shape of the opacity function.''

Using Eq. (8),
the excess ratios can also be calculated.
The median excess ratios
and standard deviations are tabulated in Tables 
\ref{table_exratio} and \ref{table_exratio2600}.
The excess ratios were calculated with 
the MSX bands, WISE W3 and W4, and GLIMPSE [8.0] bands. They appear to be described by constant values and 
can provide constraints on the circumstellar
extinction curve of the sampled O-rich AGB stars
\citep[e.g.,][]{bedijn87,volk88,ossenkopf92}.

\section{A grid of blackbody models with mass loss}
\label{grid-black}

\begin{figure*}
\begin{center}
\resizebox{0.46\hsize}{!}{\includegraphics[angle=0]{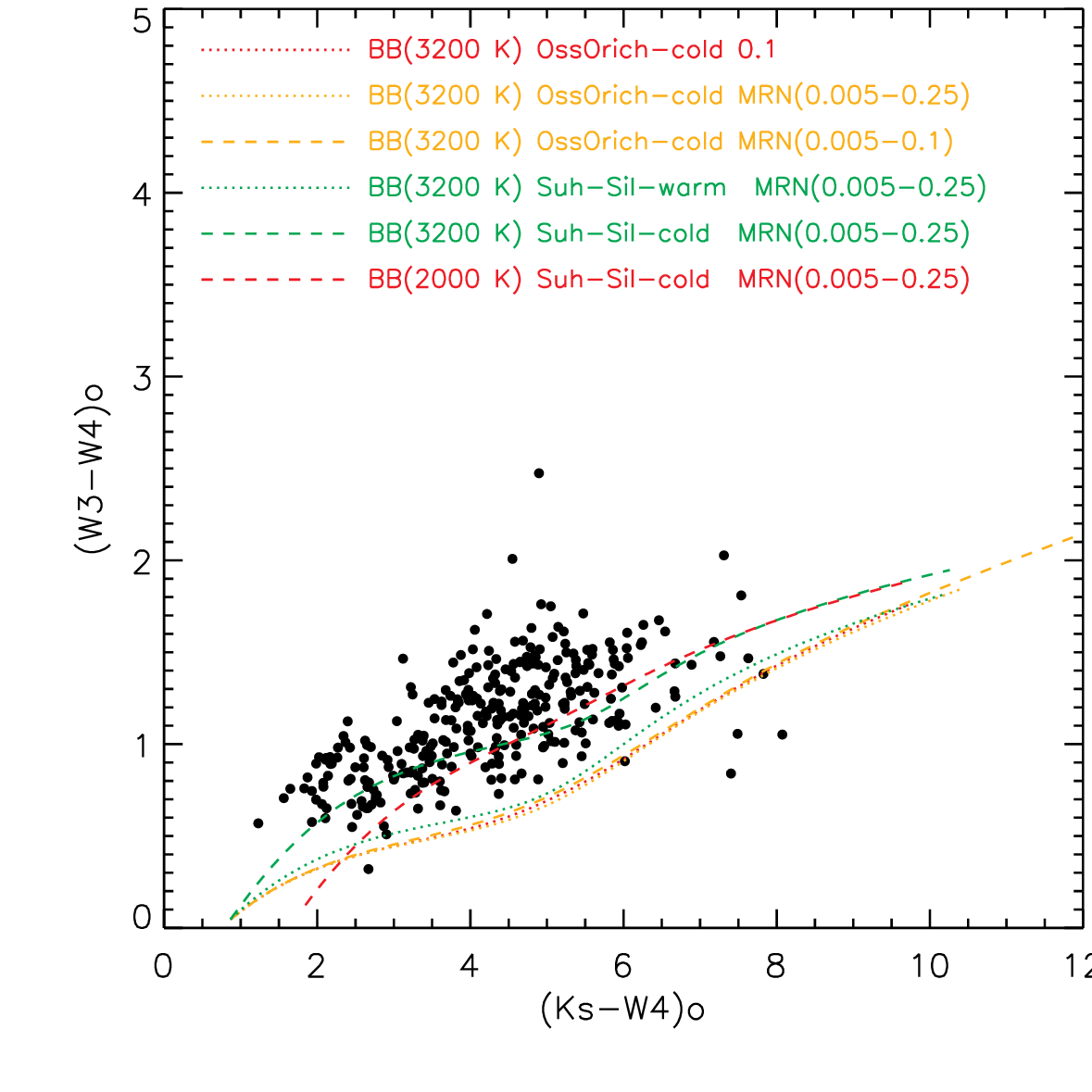}}
\resizebox{0.46\hsize}{!}{\includegraphics[angle=0]{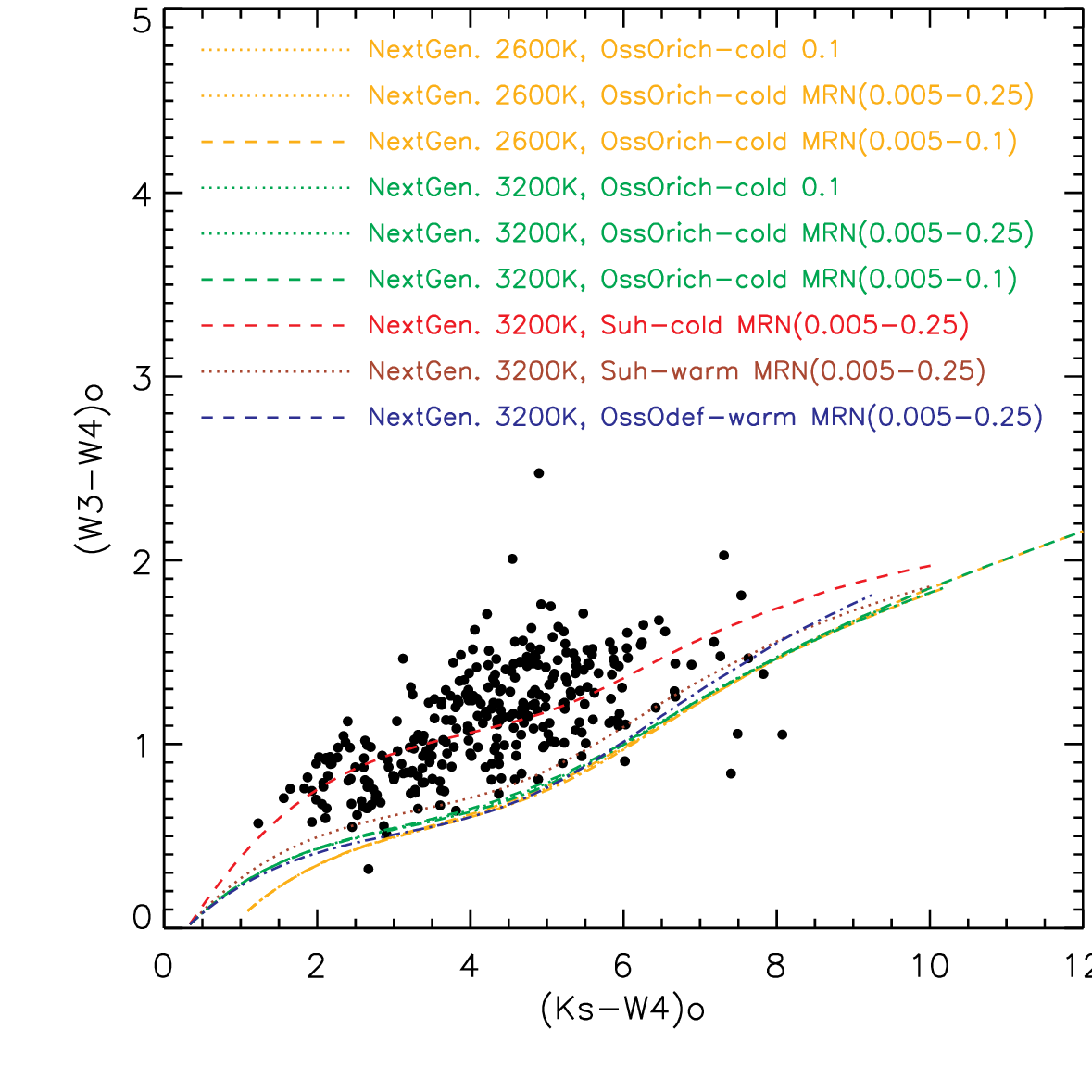}}
\resizebox{0.46\hsize}{!}{\includegraphics[angle=0]{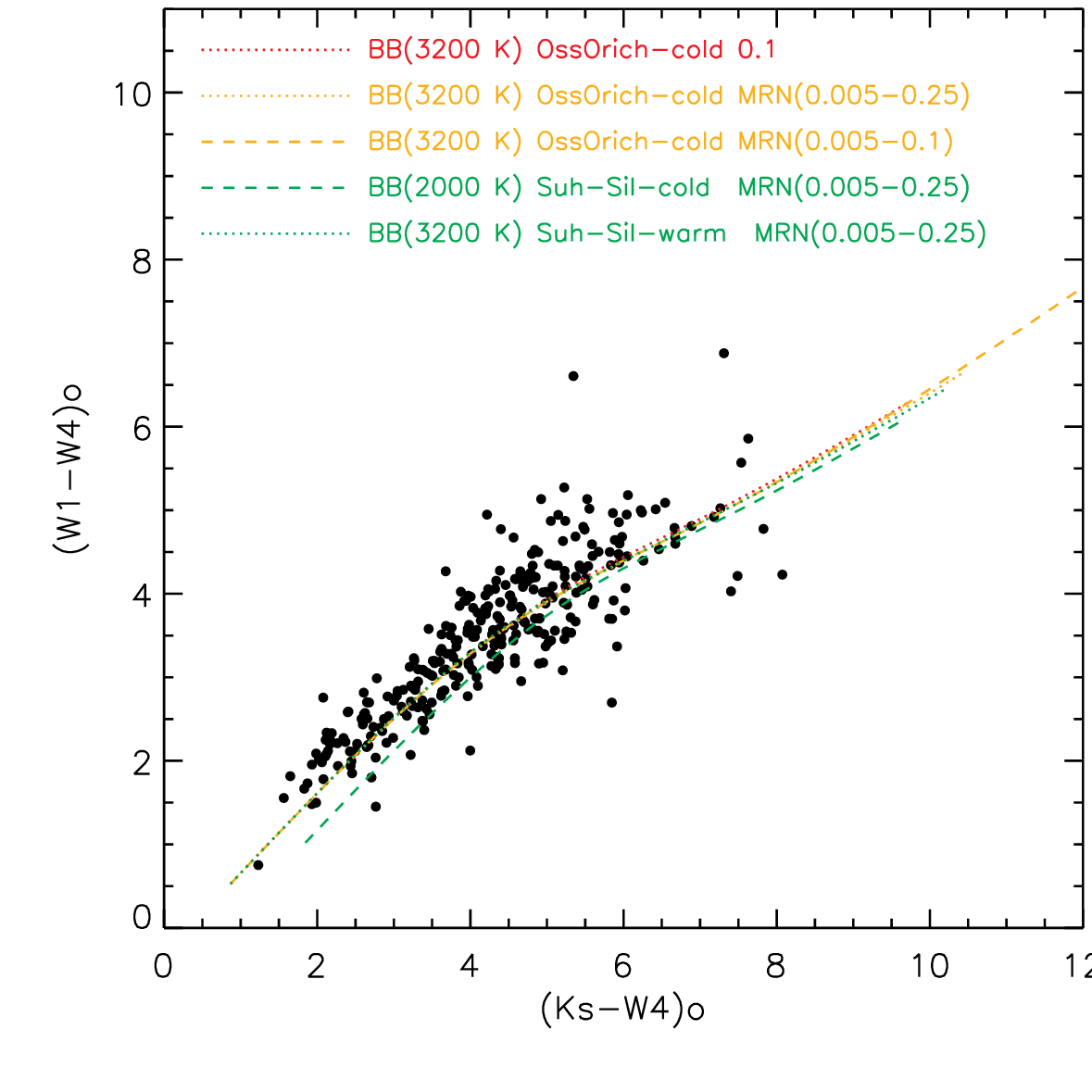}}
\resizebox{0.46\hsize}{!}{\includegraphics[angle=0]{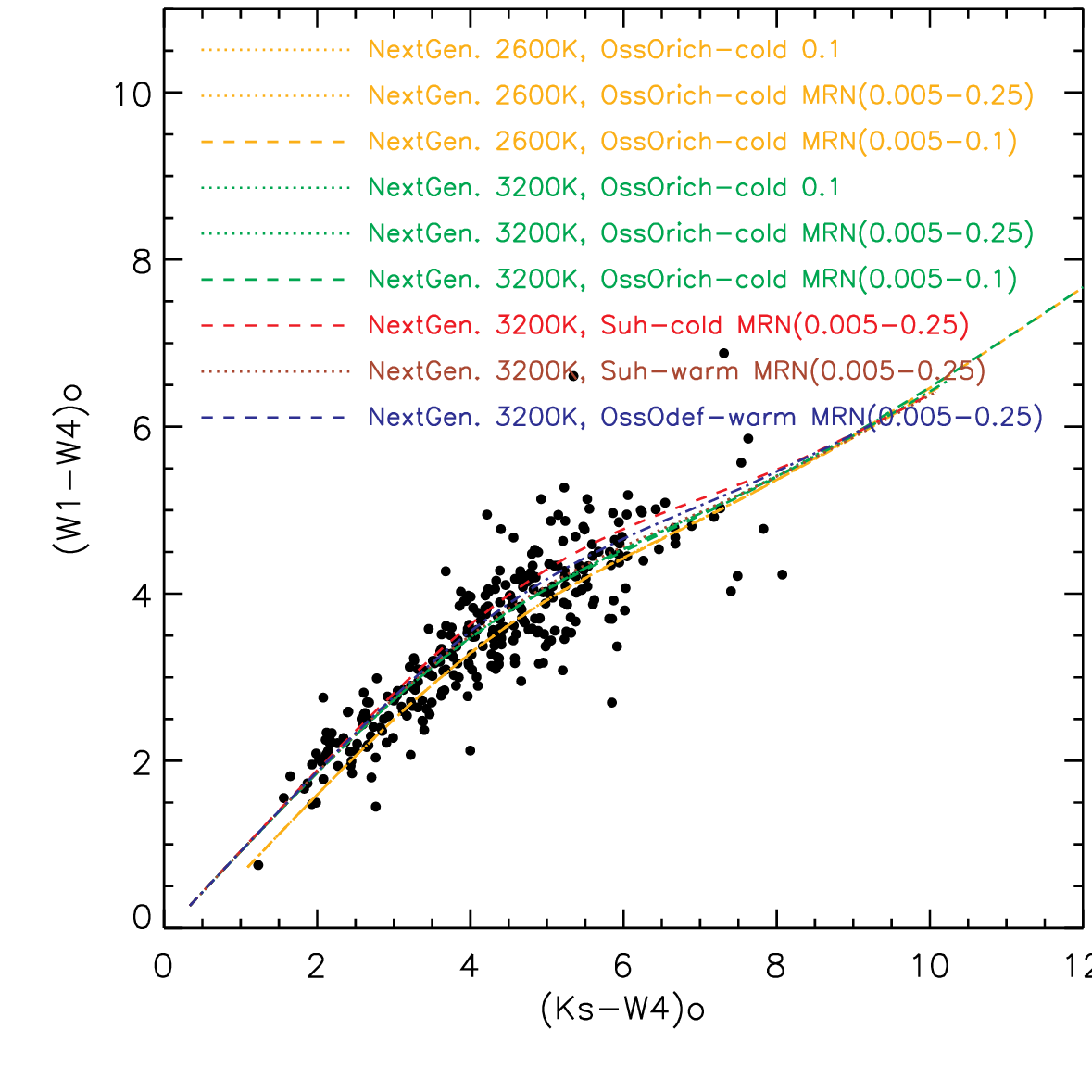}}
\end{center}
\caption{ \label{colcol_dusty} 
Dereddened (W3-W4)$_o$ vs. (\Ks-W4)$_o$ (top).
Dereddened (W1-W4)$_o$ vs. (\Ks-W4)$_o$ (bottom).
The panels on the left show the models obtained with 
 blackbodies of 3200 K  and 2000 K and the astronomical silicate of 
\citet{ossenkopf92}  and \citet{suh99}.  The panels on the right show
the models with the synthetic spectra of 3200 K and 2600 K of \citet{allard11}
and the astronomical silicate of \citet{ossenkopf92}  and \citet{suh99}.
It appears that the curves of  stars of 3200 K and 2600 K are 
similar and that the main parameter is the  dust type and the maximum grain size.
A smaller maximum grain size increases the span of colours 
(extending the curve to redder colours).
}
\end{figure*}

\begin{figure}
\begin{center}
\resizebox{0.8\hsize}{!}{\includegraphics[angle=0]{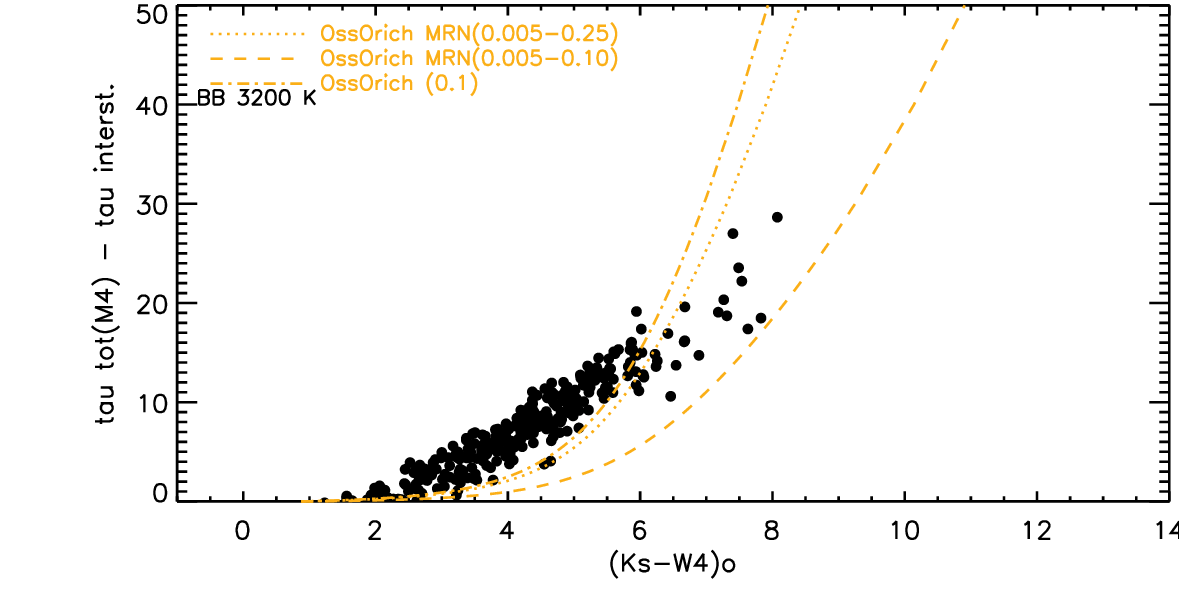}}
\resizebox{0.8\hsize}{!}{\includegraphics[angle=0]{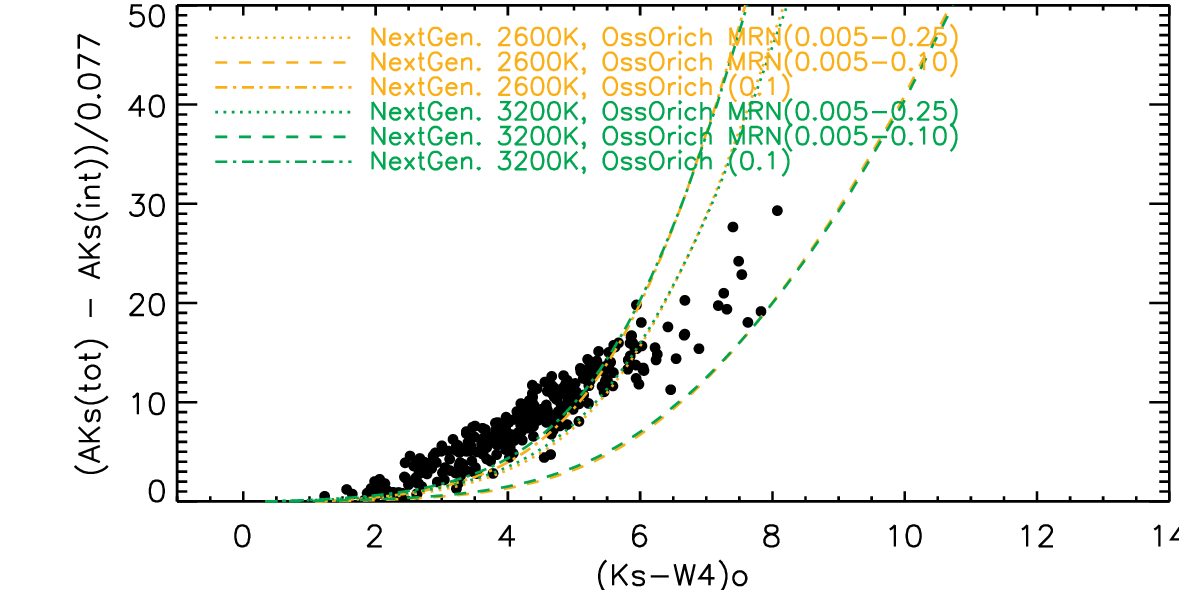}}
\end{center}
\caption{\label{coltau_dusty}
 $\tau$ envelope ($\tau$ total minus $\tau$ interstellar) values
of stars in \citet{messineo18} 
are plotted versus the dereddened (\Ks-W4)$_o$ colours.
{\it Top panel}: Dusty models with a blackbody of 3200 K.{\  Bottom panel} shows  dusty models 
of giant synthetic spectra with \Teff\ of 2600 K (in orange) and 3200 K
(in green), logg=1, and Z=0 dex,
from the NextGen  library  of \citet{hauschildt99}.
For every synthetic spectrum, three different curves obtained with DUSTY
are over-plotted. 
The  dashed-dotted model
shows  dust grains with a fixed size of 0.1 \um.
The  dotted model uses the MNR distribution with a 
minimum of 0.005 \um\ and a maximum of 0.25 \um.
The dashed model is based on the MNR distribution and
has a maximum of 0.1 \um.
}
\end{figure}

\begin{figure*}
\begin{center}
\resizebox{0.45\hsize}{!}{\includegraphics[angle=0]{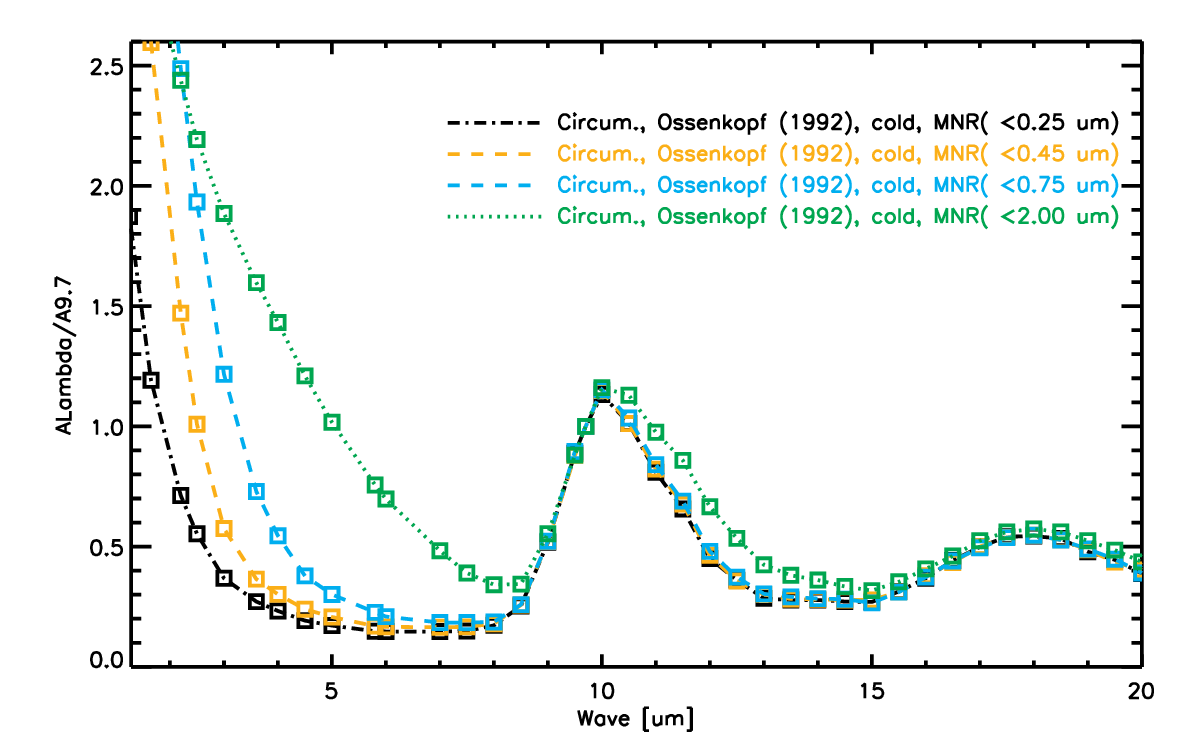}}
\resizebox{0.45\hsize}{!}{\includegraphics[angle=0]{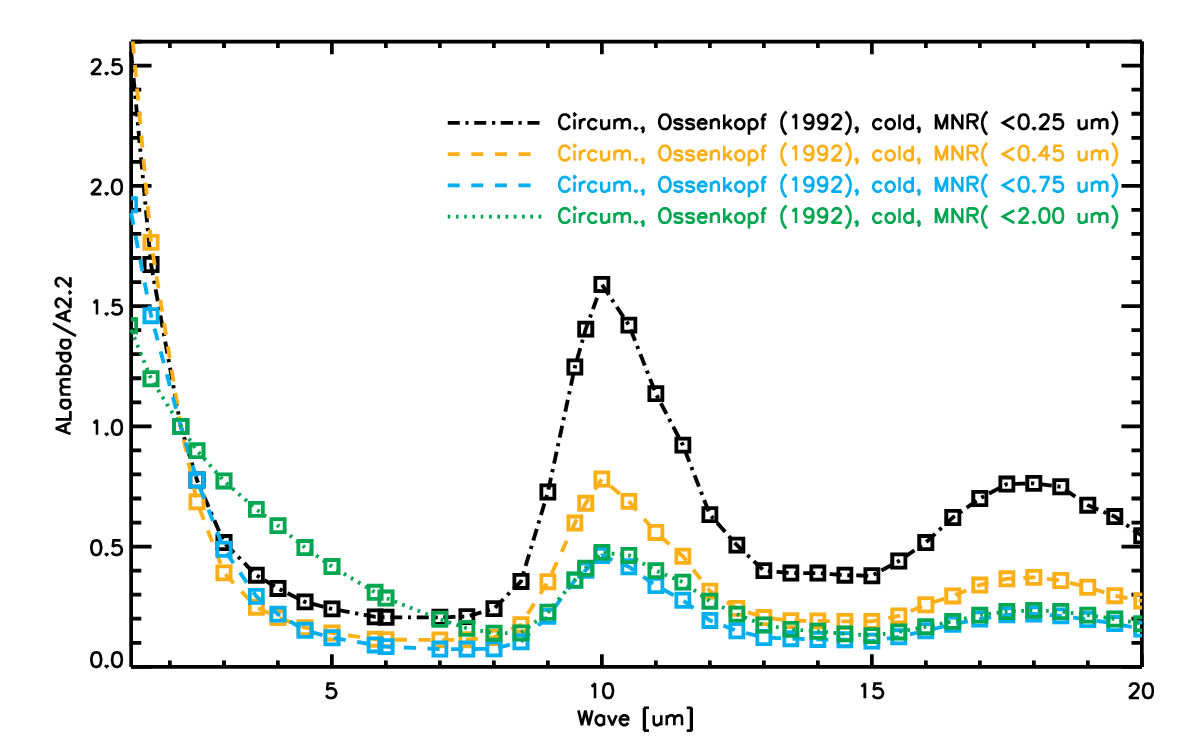}}
\end{center}
\begin{center}
\resizebox{0.45\hsize}{!}{\includegraphics[angle=0]{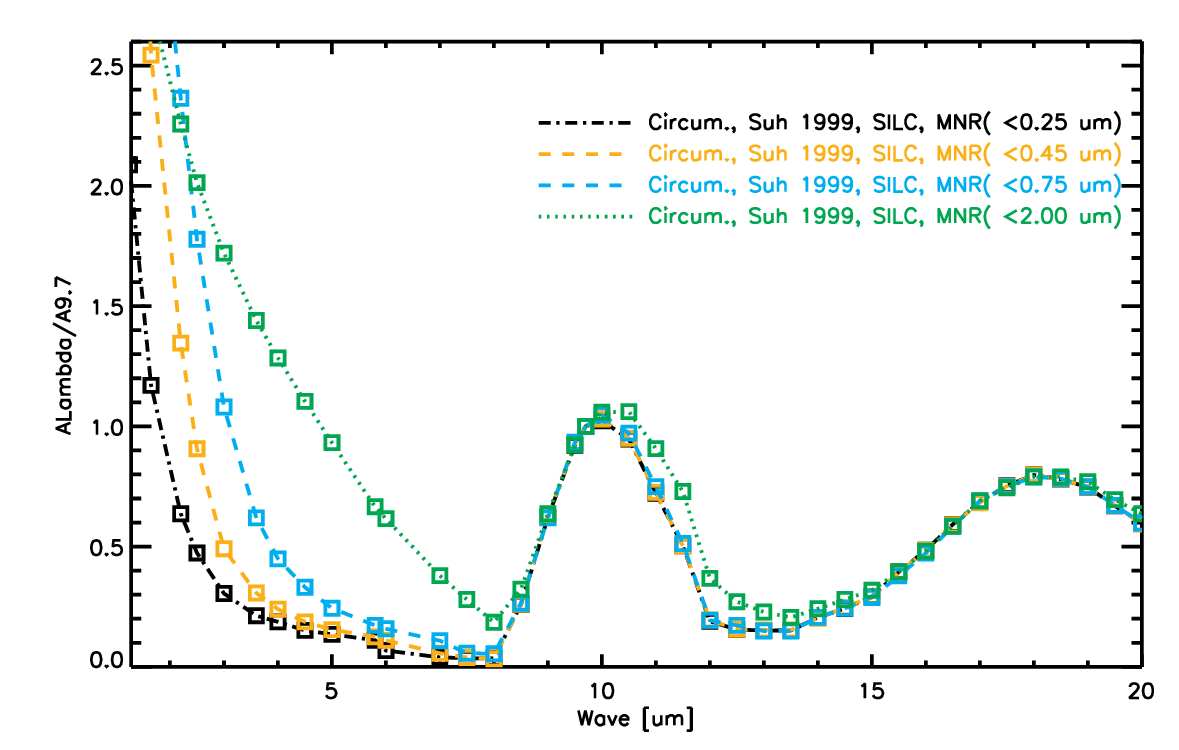}}
\resizebox{0.45\hsize}{!}{\includegraphics[angle=0]{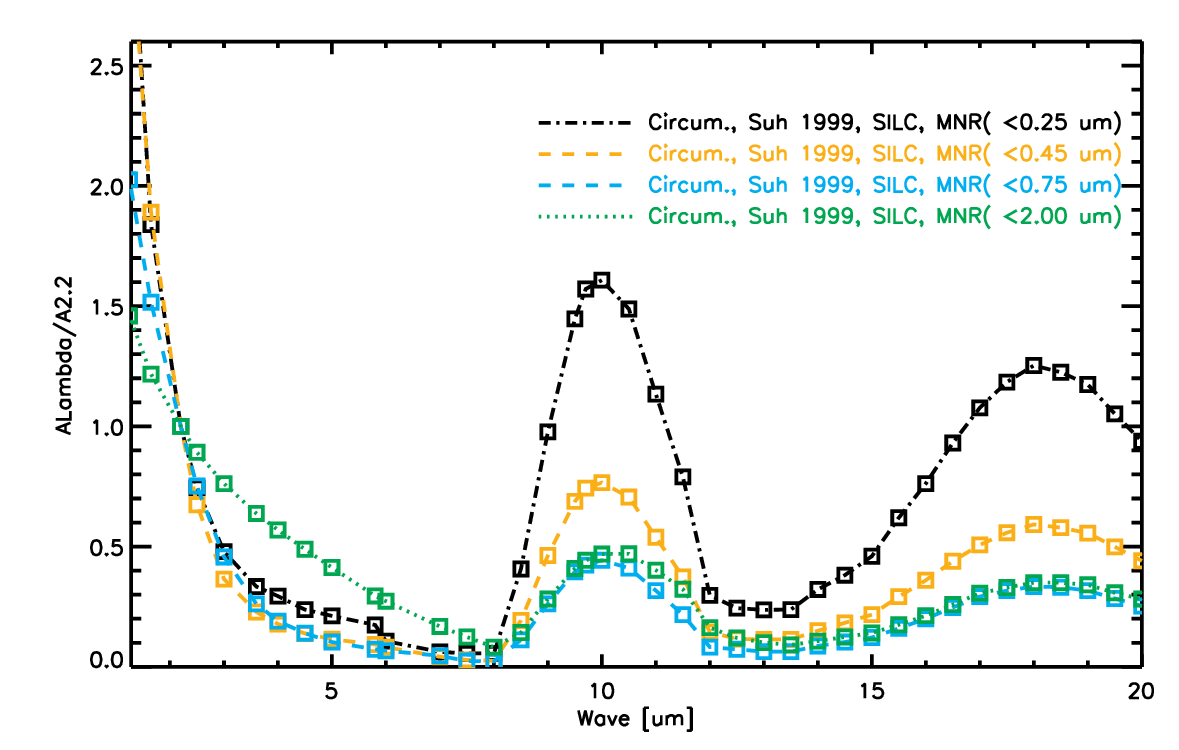}}
\end{center}
\caption{\label{fig:maxcurve}
 Four circumstellar extinction curves were estimated with DUSTY
using an MNR distribution and dust sizes from  0.005 to 0.25,
0.45,  0.75,  and 2.00 \um\ respectively. The curves are normalized to 9.7 \um\ 
in the left panels and  2.12 in the right panels.
 The astronomical silicate grains by  \citet{ossenkopf92} for a maximum grain 
size of 0.25 \um\ (black), 0.45\um\ (orange), 0.75\um\ (cyan), and 2.00\um\ (green)
are used in the two upper panels;
while in the two lower panels the silicate grains of \citet{suh99} are used.
The squares mark the circumstellar effective extinction ratios
measured on SED models generated with the DUSTY code. The curves are obtained by
interpolating the ratios to a finer grid (0.1 \um).
}
\end{figure*}

\begin{table*}
\caption{\label{table_exratiotheory}   Circumstellar model excess ratios}
\begin{tabular}{ll|lll|lll|lllll}
\hline
\hline
Excess ratios &                  Temp& \multicolumn{3}{c}{ Ossenkopf et al. (1992)-O-richCold }  &\multicolumn{3}{c}{ Suh (1999)-SILC }  & \multicolumn{3}{c}{ Rosenthal et al. (2000) } \\
              &                      &   curve\_1  & curve\_2  & curve\_3 &curve\_4  & curve\_5  & curve\_6     &curve\_7 &curve\_8&curve\_9  \\
              &                  [K]&  (max\_0.25) & (max\_0.75)&(max\_2.00)&(max\_0.25)& (max\_0.75)&(max\_2.00)& (peak\_1.4)&(peak\_0.4)& (peak\_0.2)\\
\hline
\hline
 $\frac{E(Ks-A)}{E(Ks-D)}$        &3200&   0.990 &  0.964 &  0.895 &  1.144 &  0.979 &  0.899 &    0.889 &  0.882 &  0.905 &  \\
 $\frac{E(Ks-C)}{E(Ks-D)}$        &3200&   0.624 &  0.905 &  0.854 &  0.966 &  0.979 &  0.923 &    0.548 &  0.906 &  0.956 &  \\
 $\frac{E(Ks-E)}{E(Ks-D)}$        &3200&   0.871 &  0.971 &  0.983 &  0.417 &  0.880 &  0.869 &    0.715 &  0.967 &  1.000 &  \\
 $\frac{E(Ks-W1)}{E(Ks-W3)}$      &3200&   1.321 &  0.790 &  0.404 &  1.219 &  0.806 &  0.413 &    1.738 &  0.693 &  0.608 &  \\
 $\frac{E(Ks-W2)}{E(Ks-W3)}$      &3200&   1.674 &  1.046 &  0.677 &  1.499 &  1.040 &  0.671 &    2.368 &  0.943 &  0.828 &  \\
 $\frac{E(Ks-W4)}{E(Ks-W3)}$      &3200&   1.367 &  1.089 &  1.132 &  0.687 &  0.983 &  1.006 &    1.467 &  1.121 &  1.078 &  \\
 $\frac{E(Ks-[3.6])}{E(Ks-[5.8])}$&3200&   0.779 &  0.759 &  0.504 &  0.781 &  0.779 &  0.519 &    0.701 &  0.701 &  0.701 &  \\
 $\frac{E(Ks-[4.5])}{E(Ks-[5.8])}$&3200&   0.921 &  0.923 &  0.733 &  0.898 &  0.925 &  0.734 &    0.872 &  0.872 &  0.872 &  \\
 $\frac{E(Ks-[8.0])}{E(Ks-[5.8])}$&3200&   0.926 &  1.002 &  1.210 &  0.991 &  1.009 &  1.219 &    0.806 &  1.006 &  1.060 &  \\
\hline
$\frac{E(Ks-A)}{E(Ks-D)}$        &2600&   0.969 &  0.958 &  0.891 &  1.125 &  0.972 &  0.894 &    0.858 &  0.875 &  0.903 &  \\
$\frac{E(Ks-C)}{E(Ks-D)}$        &2600&   0.614 &  0.905 &  0.853 &  0.969 &  0.979 &  0.923 &    0.539 &  0.905 &  0.956 &  \\
$\frac{E(Ks-E)}{E(Ks-D)}$        &2600&   0.869 &  0.971 &  0.982 &  0.400 &  0.878 &  0.869 &    0.709 &  0.968 &  1.001 &  \\
$\frac{E(Ks-W1)}{E(Ks-W3)}$      &2600&   1.365 &  0.799 &  0.408 &  1.257 &  0.815 &  0.416 &    1.798 &  0.695 &  0.608 &  \\
$\frac{E(Ks-W2)}{E(Ks-W3)}$      &2600&   1.726 &  1.051 &  0.680 &  1.539 &  1.043 &  0.672 &    2.442 &  0.944 &  0.826 &  \\
$\frac{E(Ks-W4)}{E(Ks-W3)}$      &2600&   1.398 &  1.093 &  1.138 &  0.685 &  0.986 &  1.009 &    1.490 &  1.123 &  1.078 &  \\
$\frac{E(Ks-[3.6])}{E(Ks-[5.8])}$&2600&   0.781 &  0.763 &  0.507 &  0.781 &  0.782 &  0.521 &    0.703 &  0.703 &  0.703 &  \\
$\frac{E(Ks-[4.5])}{E(Ks-[5.8])}$&2600&   0.920 &  0.924 &  0.734 &  0.898 &  0.926 &  0.736 &    0.873 &  0.873 &  0.873 &  \\
$\frac{E(Ks-[8.0])}{E(Ks-[5.8])}$&2600&   0.920 &  1.001 &  1.212 &  0.985 &  1.008 &  1.223 &    0.793 &  1.004 &  1.060 &  \\
\hline
\end{tabular} 
\begin{list}{}
\item { Circumstellar excess ratios were calculated with
the MSX, WISE, and GLIMPSE filter response curves, 
with a stellar spectrum \citep{allard11} and  an extinction curve. 
 The columns  curve\_1, curve\_2, and curve\_3 
refer to the circumstellar extinction curves  obtained with  DUSTY and the cold O-rich
silicates by \citet{ossenkopf92} adopting a maximum size of 
0.25, 0.75, and 2.0  \um, respectively (see text). 
The columns  curve\_4, curve\_5, and  curve\_6 
refer to the circumstellar extinction curves  obtained with  DUSTY and 
the cold silicates (SILC) by \citet{suh99} adopting a maximum size of 
0.25, 0.75, and 2.0  \um, respectively, (see text). 
The spectrum of a naked star of 3200 K is used
to obtain the naked star colours in the upper part of the table (first 9 rows).
In the lower part of the table, the spectrum of a naked star of 2600 K is used.
For comparison, the columns curve\_7, curve\_8, and curve\_9  
refer to the parametric formula for Galactic interstellar extinction by 
\citet[][]{rosenthal00} normalized at 2.12 \um\
with the silicate peak set at 1.4, 0.4, and 0.2, respectively.
}
\end{list}
\end{table*}

We assumed a fixed temperature of 3200 K (M6),
the astronomical silicate grains
of  \citet{ossenkopf92} 
as well as the silicate grains of \citet{suh99}. We used the DUSTY code of \citet{ivezic99}
to generate a set of SED models with  
optical depth,  $\tau _{0.55{\rm \mu m}}$, from 0 to 40.
The  model consists of a spherical shell 
with density decreasing with   $R^{-2}$.
The dust condensation radius was set to 1000 K. 

In the diagram depicting $W1-W4$ versus \Ks$-W4$, the  models aptly
describe the  data points of \citet{messineo18}, 
as shown in Fig. \ref{colcol_dusty}.
In the diagram $W3-W4$ versus \Ks$-W4$, the models  with the Ossenkopf's
silicate grains display smaller  $W3-W4$ than the data points,
 while the models with the cold silicates by \citet{suh99}
reproduce the observed colours \citep{suh21}. Indeed, the W3
filter encompasses the 10 \um, as well as the 
  13 \um, and the 18 \um\ dust features.
 SED models with  Suh's silicates
and models with  Ossenkopf's silicates
agree   short-ward of 10 \um; but they disagree
around 18 \um\ \citep{suh99}.
Suh's warm silicates generate a  smaller  emission feature at 18 \um\ than 
the cold silicates \citep{suh99}. Suh's cold silicates  
generate a stronger absorption feature at 18 \um\ and a
stronger continuum longward of 18 \um\
than Ossenkopf's silicates \citep{suh99} and generate larger W3-W4 colours.

Figure \ref{colcol_dusty} shows how changing
the $\tau$ results in a colour-colour track,
given a stellar temperature and a dust  type.
Each point on the track represents a specific tau-point.
However, when varying the dust size, the $\tau$\ scale changes.
This can be seen in Fig. \ref{coltau_dusty},
where the dereddened (\Ks$-W4)_o$ is plotted against 
the $\tau$\ envelope.
A larger maximum size results in a larger $\tau$
for a given colour.
In recent literature,  
the standard distribution of grain sizes, MNR, is usually adopted with 
maximum sizes from 0.1 to 1. \um\ 
 \citep[e.g.,][]{olofsson22,vanloon05,wang21} 
or fixed sizes of  0.1-0.5 \um\ \citep[e.g.,][]{blommaert18,suh21}.

 Circumstellar effective extinction ratios are highly dependent 
on the maximum grain size.
We estimated that $\tau_{2.2}=2.48 \times \tau_{9.7}$
using the DUSTY code, the 
astronomical silicate grains by  \citet{ossenkopf92}, and an MNR
distribution from 0.005 to 0.75 \um.
For a maximum size of
0.45 \um,
$\tau_{2.2}=1.47 \times \tau_{9.7}$, 
and $\tau_{2.2}=0.71 \times \tau_{9.7}$  
when the maximum size is 0.25 \um.
Four circumstellar extinction curves  with
the MNR distribution and maximum sizes of 2.00, 0.75, 0.45, and 0.25 \um\ 
are plotted in Fig. \ref{fig:maxcurve} to illustrate this.
 The curves show different peak values of the 9.7 feature
 (0.49-0.50, 0.44-0.46, 0.77-0.78, and 1.59-1.61, respectively)
when normalizing at 2.12 \um.
Compared to a curve with a maximum grain size of 0.25 \um, 
an extinction curve with a maximum grain size of 0.75 \um\ appears 
to have a lower 9.7 \um\ peak opacity.
In the near-infrared, the same curve appears  located above
the curve  with a maximum grain size of 0.25 \um,
 when normalized at 9.7 \um.

In Table \ref{table_exratiotheory}, theoretical excess ratios 
are calculated 
by convolving the curves with a maximum grain size of  2.00 \um, 0.75 \um,\ and 0.25 \um\
with the filter profiles and a cool spectrum by \citet{allard11}.
A comparison of the theoretical excess ratios of 
Table \ref{table_exratiotheory} 
with those observationally measured  (Table \ref{table_exratio}),
suggests that an MNR distribution with a maximum grain size of 0.25 \um\
produces excess ratios that   are incompatible  with those observed
($\frac{E(Ks-C)}{E(Ks-D)}$ much lower than unity and 
$\frac{E(Ks-W2)}{E(Ks-W3)}$ above unity).
An MNR distribution with a maximum grain size of 0.75 \um\ 
provides a better agreement with the data  for the MSX data.
An MNR distribution with a maximum grain size of 2.00 \um\ 
also reproduces  the trends seen in Table \ref{table_exratio}
for the WISE and GLIMPSE filters short-ward of 8 \um\ (these 
filters are affected by molecular absorption).

In order to reproduce the observed colours,
small values of the maximum 
grain sizes (0.1-0.25 \um) are frequently reported in 
the literature. For instance, \citet{volk88}
estimated 
$\tau_{2.2}=0.4 \times \tau_{9.7}$ for their sample of IRAS sources
using dust grain sizes smaller than 0.25 \um.
However, as  \citet{david90} points out, 
observational data (IRAS data) only constrains
their opacity curve
in the mid-infrared region;
for the near-infrared curve, the theoretical 
model by \citet{draine85} is used.
\citet{bedijn87}
uses a power law with an index $-1.5$
for the opacity curve below 5 \um,  yielding 
$\tau_{1.65}=0.67 \times \tau_{9.7}$.

Recently, \citet{maercker22} preferred larger grain sizes (2 \um) to
fit the far-infrared fluxes of Carbon AGB stars.
Large dust grains with an average size of 0.5 \um\ 
were detected in the envelopes of VY Canis Majoris \citep{scicluna15}
and of  W Hydrae \citep{ohnaka16} with optical polarimetric imaging.

\section{Discussion}
\label{verifica_maser}
Relations between the de-reddened colours and the extinction-free 
colours
Q$_{\rm_1}$, Q$_{\rm_2}$,Q$_{\rm_A}$,Q$_{\rm C}$,..Q$_{\rm W4}$
were constructed using   a set of O-rich 
Mira-like stars with known interstellar extinction
\citep{messineo05}. 
Such equations allow us to 
determine the interstellar extinction
for every Mira-like star that has accessible 
near- and mid-infrared (NIR-MIR) colours. 

In order to verify  the applicability and reliability of the new
methodology in determining interstellar extinction, 
we carried out an analysis of  existing  catalogs of Miras
with available $JHK$ and MIR\ measurements.

The most obscured Miras often exhibit  OH maser emission and 
are therefore called OH/IR stars.  
Typically, their flux distribution is modeled with radiative transfer 
codes, and their interstellar extinction  is derived from  
infrared extinction maps. 
For example, \citet{olofsson22}   analyzed a sample of  22
AGB stars with OH masers near the Galactic center 
by using CO line observations taken with   
 The Atacama large (sub-)millimeter array (Alma) array.
The map of extinction by \citet{gonzalez18} 
and the extinction law by \citet{nishiyama09} were used.
Archival $JHK$ photometry 
(from the 2MASS, VVV, and UKIDSS  
surveys\footnote{VVV stands for the VISTA Variables in the Via
Lactea survey  \citep{soto13}. The UKIDSS Galactic Plane Survey (GPS)
is described in \citet{lucas08}.}, the catalogues of 
\citet{nogueras19} and \citet{nogueras21}) and 
mid-infrared photometry from   GLIMPSE surveys (5.8 and 8.0 um) 
could be retrieved from the VIZIER database for 8 stars of the
Olofsson's sample.
The equations of Table \ref{table.qxfits} were used
to make interstellar extinction estimates  
(\Aks(int) from 0.90 to 2.0 mag).
The two sets of \Aks(int) linearly correlate 
with a standard deviation $\sigma$=0.49 mag and
a mean difference  (this work - Olofsson's values)
of 0.09 mag (or 0.24 mag when rescaling the extinction 
to a power law with index = $-$2.1).
The non-coevality of the near- and mid-infrared photometry 
accounts for the large $\sigma$; the OH/IR stars have typical 
\Ks\ amplitudes ranging from 0.9 to 2.5 mag \citep{messineo20}.

Using data from the ISOGAL survey \citep{omont03}, 
\citet{ojha03} determined interstellar extinction, mass loss rate,
and luminosity values for 321   late-type stars located
in  Bulge fields at ($|l|<2^\circ$, $|b| \approx 1^\circ-4^\circ$). 
The majority of the sampled stars are AGB stars because 
they are brighter than the stars at the tip of the red giant branch.
The authors used DENIS $IJKs$ and ISOGAL photometry at 7 and 15 \um\ 
and the silicate models by \citet{jeong03}.
The new method was used to estimate \Ak(int) for 228 stars 
in this sample based on GLIMPSE and 2MASS data.
The Ojha's A$_V$ values are converted to \Aks\ 
using a factor of 0.089 \citep{ojha03}
and then\ compared with the \Aks\  obtained here.
The  mean difference is 0.09 mag and a $\sigma=0.20$ mag.

\citet{matsunaga09} estimate the  \Ak(int) values
of 52 Mira stars at the Galactic center
using average near-infrared measurements 
and the extinction law by \citet{nishiyama06}.
The average $JHK$ measurements along with the 
GLIMPSE [5.8] and [8.0] magnitudes are used to compute 
the \Aks(int) values for 30 stars
using the new method. 
The  \Aks(HK) from Matsunaga  were converted  
to a power law with index =$-2.1$ (by multiplying for 0.91) 
and range from 1.81 to 2.59 mag. 
An average  difference 
between Matsunaga's extinction values and the here estimated values
of +0.1 mag is obtained and  $\sigma=0.16$ mag. 

The established equations 
between intrinsic infrared colours and extinction-free
colours (Q$_\lambda$), which are listed in Table \ref{table.qxfits},
are useful to automatically determine stellar obscuration. 
This set of equations 
can be used to select  late-type stars in a given sky region,
and to improve their estimates of luminosity.

Assuming a reasonable estimate of the total, 
the envelope extinction can be calculated as the difference 
between the total and the interstellar extinction.
Mathematically, it is a straightforward calculation.
In reality, estimating the circumstellar extinction 
is actually difficult because accurate spectral types are lacking, 
naked-star colours are unknown
and affected by strong variable molecular absorption.

Future photometric multi-wavelengths and multi-epochs
surveys and spectroscopic surveys,
such as LSST, Gaia DR4, 4MOST, and GALAH, 
will enable us to obtain spectral types 
and temperatures, and their variations.
As a result, we will be able to obtain   more accurate extinction calculations.

\begin{acknowledgements}

This work has made use of data from the European Space Agency (ESA) mission {\it Gaia}
($http://www.cosmos.esa.int/gaia$), processed by the {\it Gaia} Data Processing and Analysis
Consortium (DPAC, $http://www.cosmos.esa.int/web/gaia/dpac/consortium$). Funding for the DPAC
has been provided by national institutions, in particular the institutions participating in the {\it
Gaia} Multilateral Agreement. 
This publication makes use of data products from the Two Micron All
Sky Survey, which is a joint project of the University of Massachusetts and the Infrared Processing
and Analysis Center / California Institute of Technology, funded by the National Aeronautics and Space
Administration and the National Science Foundation. 
This work is based on observations made with
the Spitzer Space Telescope, which is operated by the Jet Propulsion Laboratory, California
Institute of Technology under a contract with NASA.
This research made use of data products from the Midcourse Space Experiment, the processing of which
was funded by the Ballistic Missile Defense Organization with additional support from the NASA
office of Space Science. 
This publication makes use of data products from WISE, which is a joint
project of the University of California, Los Angeles, and the Jet Propulsion Laboratory / California
Institute of Technology, funded by the National Aeronautics and Space Administration. 
This research
has made use of the VizieR catalogue access tool, CDS, Strasbourg, France, and SIMBAD database.
This research utilized  the NASA’s Astrophysics Data System Bibliographic Services. 
This work is based on the PhD thesis by Messineo M.  (2004) which was
supported by the Netherlands
Research School for Astronomy (NOVA) through a netwerk 2, Ph.D.
stipend.
MM thanks  Dr. Harm Habing  and Dr. Frank Bertoldi for 
insightful discussions on interstellar extinction during her PhD thesis, 
and  the anonymous referee for his constructive inputs.
\end{acknowledgements}


\newpage
\begin{appendix}
\section{$Q_\lambda$ and the jolly equation}

This is the definition of (Ks-$\lambda$) colour excess:
\begin{equation}
E(Ks-\lambda)= A_{Ks} \times ( 1 - \frac{ A_\lambda}{ A_{Ks}}) 
,\end{equation}

\begin{equation}
A_{Ks} {\rm (interstellar)} =
\frac{[(Ks-\lambda)-(Ks-\lambda)o] }{ ( 1 - \frac{ A_\lambda}{ A_{Ks}})}
,\end{equation}

\begin{equation}
A_{Ks} {\rm (interstellar)} =
\frac{[(H-Ks)-(H-Ks)o] }{  \frac{ A_H }{(A_{Ks}} - 1 ) }
,\end{equation}

By combining  Eq. A.2 and A.3, it is obtained
\begin{equation}
\frac{[(Ks-\lambda)-(Ks-\lambda)o]} { ( 1 - \frac{ A_\lambda}{ A_{Ks}})} = 
\frac{ [(H-Ks)-(H-Ks)o] }{ ( \frac{ A_H }{A_{Ks}} - 1 )}
,\end{equation}

 therefore,
\begin{equation}
(H-Ks)-(Ks-\lambda) *  Co  = 
(H-Ks)o-(Ks-\lambda)o * Co = Q_\lambda
,\end{equation}

where 
\begin{equation}
Co=  \frac{( \frac{ A_H }{A_{Ks}} - 1 ) }{ ( 1 -\frac{ A_\lambda}{ A_{Ks}})}   
.\end{equation}

The A.5 is the definition of the $Q_\lambda^{HK}$ parameter.
The first member contains observed quantities,
while the second term only contains intrinsic colours.
This equivalence proves that $Q_\lambda^{HK}$ does not depend
on interstellar extinction. It is an intrinsic colour of the star. Analogously, a function $Q_\lambda^{JK}$ can be defined with the $J-Ks$ colour.

\section{BC and intrinsic colours versus \Ql.}
In \citet{messineothesis}, \BCKs\ values as a function of 
infrared colours were obtained for the MSX filters.
\Mbol\ values and dereddened WISE and GLIMPSE magnitudes from 
the revised  catalog of \citet{messineo18} 
are here used to  extend the BC computation to the WISE and Glimpse filters.
Polynomial fits are listed in Table \ref{table.fit}.

\begin{table*}
\caption{\label{table.fit} Polynomial fits to the infrared BC values 
versus the dereddened colours.}
\begin{tabular}{llrrrrrr}
\hline
\hline
X-axis        & Y-axis       & Coef~n1               & Coef~n2               & Coef~n3          &  $<$Y-axis $-$ Y-Fit$>$ & $\sigma$ &X-range   \\
(mag)         & (mag)        &\\
\hline
      (Ks$-$A)o   &      BC(Ks)    &  2.263$\pm$  0.039    &  0.691$\pm$  0.028    & -0.156$\pm$  0.005    & -0.000    &  0.210 &$[0.5,7.0]$  \\
      (Ks$-$A)o   &       BC(A)    &  2.263$\pm$  0.039    &  1.691$\pm$  0.028    & -0.156$\pm$  0.005    &  0.000    &  0.210 &$[0.5,7.0]$  \\
      (Ks$-$C)o   &      BC(Ks)    &  1.594$\pm$  0.069    &  0.882$\pm$  0.036    & -0.136$\pm$  0.004    & -0.000    &  0.224 &$[1.0,8.0]$  \\
      (Ks$-$C)o   &       BC(C)    &  1.594$\pm$  0.069    &  1.882$\pm$  0.036    & -0.136$\pm$  0.004    & -0.000    &  0.224 &$[1.0,8.0]$  \\
      (Ks$-$D)o   &      BC(Ks)    &  1.618$\pm$  0.068    &  0.861$\pm$  0.036    & -0.131$\pm$  0.004    &  0.000    &  0.223 &$[1.0,8.0]$  \\
      (Ks$-$D)o   &       BC(D)    &  1.618$\pm$  0.068    &  1.861$\pm$  0.036    & -0.131$\pm$  0.004    &  0.000    &  0.223 &$[1.0,8.0]$  \\
      (Ks$-$E)o   &      BC(Ks)    &  1.391$\pm$  0.131    &  0.852$\pm$  0.057    & -0.111$\pm$  0.006    &  0.000    &  0.230 &$[2.0,9.0]$  \\
      (Ks$-$E)o   &       BC(E)    &  1.391$\pm$  0.131    &  1.852$\pm$  0.057    & -0.111$\pm$  0.006    &  0.000    &  0.230 &$[2.0,9.0]$  \\
\hline
     (Ks$-$W1)o    &      BC(Ks)    &  2.943$\pm$  0.028    &  0.172$\pm$  0.042    & -0.245$\pm$  0.013    &  0.000    &  0.285 &$[0.0,4.0]$  \\
     (Ks$-$W1)o    &      BC(W1)    &  2.943$\pm$  0.028    &  1.172$\pm$  0.042    & -0.245$\pm$  0.013    &  0.000    &  0.285 &$[0.0,4.0]$  \\
     (Ks$-$W2)o    &      BC(Ks)    &  2.839$\pm$  0.032    &  0.264$\pm$  0.028    & -0.136$\pm$  0.005    &  0.000    &  0.270 &$[0.0,6.0]$  \\
     (Ks$-$W2)o    &      BC(W2)    &  2.839$\pm$  0.032    &  1.264$\pm$  0.028    & -0.136$\pm$  0.005    & -0.000    &  0.270 &$[0.0,6.0]$  \\
     (Ks$-$W3)o    &      BC(Ks)    &  2.113$\pm$  0.057    &  0.659$\pm$  0.032    & -0.121$\pm$  0.004    & -0.000    &  0.255 &$[1.0,8.0]$  \\
     (Ks$-$W3)o    &      BC(W3)    &  2.113$\pm$  0.057    &  1.659$\pm$  0.032    & -0.121$\pm$  0.004    & -0.000    &  0.255 &$[1.0,8.0]$  \\
     (Ks$-$W4)o    &      BC(Ks)    &  1.490$\pm$  0.084    &  0.799$\pm$  0.037    & -0.104$\pm$  0.004    &  0.000    &  0.271 &$[2.0,9.0]$  \\
     (Ks$-$W4)o    &      BC(W4)    &  1.490$\pm$  0.084    &  1.799$\pm$  0.037    & -0.104$\pm$  0.004    & -0.000    &  0.271 &$[2.0,9.0]$  \\
\hline
  (Ks$-[3.6]$)o    &    BC(Ks)    &  3.094$\pm$  0.021    & -0.064$\pm$  0.024    & -0.153$\pm$  0.011    &  0.000    &  0.316    &$[-2.0,5.0]$\\
  (Ks$-[3.6]$)o    &BC($[3.6]$    &  3.094$\pm$  0.021    &  0.936$\pm$  0.024    & -0.153$\pm$  0.011    & -0.000    &  0.316    &$[-2.0,5.0]$\\
  (Ks$-[4.5]$)o    &      BC(Ks)    &  2.986$\pm$  0.024    &  0.255$\pm$  0.037    & -0.180$\pm$  0.012    & -0.000    &  0.293  &$[-1.0,5.0]$ \\
  (Ks$-[4.5]$)o    & BC($[4.5]$)    &  2.986$\pm$  0.024    &  1.255$\pm$  0.037    & -0.180$\pm$  0.012    & -0.000    &  0.293  &$[-1.0,5.0]$ \\
  (Ks$-[5.8]$)o    &      BC(Ks)    &  2.637$\pm$  0.034    &  0.559$\pm$  0.034    & -0.169$\pm$  0.007    & -0.000    &  0.244  &$[0.0,6.0]$ \\
  (Ks$-[5.8]$)o    & BC($[5.8]$)    &  2.637$\pm$  0.034    &  1.559$\pm$  0.034    & -0.169$\pm$  0.007    & -0.000    &  0.244  &$[0.0,6.0]$  \\
  (Ks$-[8.0]$)o    &      BC(Ks)    &  2.490$\pm$  0.044    &  0.602$\pm$  0.036    & -0.146$\pm$  0.007    & -0.000    &  0.249  &$[0.0,6.0]$ \\
  (Ks$-[8.0]$)o    & BC($[8.0]$)    &  2.490$\pm$  0.044    &  1.602$\pm$  0.036    & -0.146$\pm$  0.007    &  0.000    &  0.249  &$[0.0,6.0]$ \\
\hline
\end{tabular}
\begin{list}{}
  \item  {\bf Note:} Y-Fit = Coef~n1 + Coef~n2 $\times$ X-axis + Coef~n3 $\times$ X-axis$^2$. \\
  The colours here used are those dereddened with the \Akint\ from surrounding stars
  by \citet{messineo05} and \citet{messineo18} 
  (rescaled to a power law with an index of $-2.1$) and the extinction Curve 3.
\end{list}
\end{table*}

\begin{table*}
\caption{\label{table.fit2} Polynomial fits to the infrared BC values 
versus the dereddened colours.}
\begin{tabular}{llrrrrrr}
\hline
\hline
X-axis        & Y-axis       & Coef~n1               & Coef~n2               & Coef~n3          &  $<$Y-axis $-$ Y-Fit$>$ & $\sigma$    \\
(mag)         & (mag)        &\\
\hline

      (Ks$-$A)o    &      BC(Ks)    &  2.540$\pm$  0.028    &  0.632$\pm$  0.021    & $-$0.156$\pm$  0.003    &  0.000    &  0.158   \\
      (Ks$-$A)o    &       BC(A)    &  2.540$\pm$  0.028    &  1.632$\pm$  0.021    & $-$0.156$\pm$  0.003    &  0.000    &  0.158   \\
      (Ks$-$C)o    &      BC(Ks)    &  1.869$\pm$  0.051    &  0.857$\pm$  0.027    & $-$0.140$\pm$  0.003    &  0.000    &  0.169   \\
      (Ks$-$C)o    &       BC(C)    &  1.869$\pm$  0.051    &  1.857$\pm$  0.027    & $-$0.140$\pm$  0.003    &  0.000    &  0.169   \\
      (Ks$-$D)o    &      BC(Ks)    &  1.834$\pm$  0.055    &  0.846$\pm$  0.028    & $-$0.132$\pm$  0.003    &  0.000    &  0.176   \\
      (Ks$-$D)o    &       BC(D)    &  1.834$\pm$  0.055    &  1.846$\pm$  0.028    & $-$0.132$\pm$  0.003    &  0.000    &  0.176   \\
      (Ks$-$E)o    &      BC(Ks)    &  1.605$\pm$  0.104    &  0.850$\pm$  0.045    & $-$0.114$\pm$  0.005    &  0.000    &  0.183   \\
      (Ks$-$E)o    &       BC(E)    &  1.605$\pm$  0.104    &  1.850$\pm$  0.045    & $-$0.114$\pm$  0.005    &  0.000    &  0.183   \\
\hline

     (Ks$-$W1)o    &      BC(Ks)    &  3.088$\pm$  0.026    &  0.154$\pm$  0.037    & $-$0.239$\pm$  0.011    &  0.000    &  0.248   \\
     (Ks$-$W1)o    &      BC(W1)    &  3.088$\pm$  0.026    &  1.154$\pm$  0.037    & $-$0.239$\pm$  0.011    &  0.000    &  0.248   \\
     (Ks$-$W2)o    &      BC(Ks)    &  3.074$\pm$  0.028    &  0.165$\pm$  0.024    & $-$0.126$\pm$  0.005    &  0.000    &  0.229   \\
     (Ks$-$W2)o    &      BC(W2)    &  3.074$\pm$  0.028    &  1.165$\pm$  0.024    & $-$0.126$\pm$  0.005    &  0.000    &  0.229   \\
     (Ks$-$W3)o    &      BC(Ks)    &  2.459$\pm$  0.047    &  0.562$\pm$  0.027    & $-$0.117$\pm$  0.004    &  0.000    &  0.218   \\
     (Ks$-$W3)o    &      BC(W3)    &  2.459$\pm$  0.047    &  1.562$\pm$  0.027    & $-$0.117$\pm$  0.004    &  0.000    &  0.218   \\
     (Ks$-$W4)o    &      BC(Ks)    &  1.789$\pm$  0.074    &  0.751$\pm$  0.032    & $-$0.103$\pm$  0.003    &  0.000    &  0.238   \\
     (Ks$-$W4)o    &      BC(W4)    &  1.789$\pm$  0.074    &  1.751$\pm$  0.032    & $-$0.103$\pm$  0.003    &  0.000    &  0.238   \\

\hline
  (Ks$-[3.6]$)o    &    BC(Ks)    &  3.270$\pm$  0.019      & $-$0.079$\pm$0.022    & $-$0.162$\pm$  0.009    &  0.000    &  0.265   \\
  (Ks$-[3.6]$)o    &BC($[3.6]$    &  3.270$\pm$  0.019      &  0.921$\pm$  0.022    & $-$0.162$\pm$  0.009    &  0.000    &  0.265   \\
  (Ks$-[4.5]$)o    &      BC(Ks)    &  3.158$\pm$  0.021    &  0.199$\pm$  0.032    & $-$0.177$\pm$  0.010    &  0.000    &  0.243   \\
  (Ks$-[4.5]$)o    & BC($[4.5]$)    &  3.158$\pm$  0.021    &  1.199$\pm$  0.032    & $-$0.177$\pm$  0.010    &  0.000    &  0.243   \\
  (Ks$-[5.8]$)o    &      BC(Ks)    &  2.884$\pm$  0.027    &  0.462$\pm$  0.027    & $-$0.162$\pm$  0.006    &  0.000    &  0.198   \\
  (Ks$-[5.8]$)o    & BC($[5.8]$)    &  2.884$\pm$  0.027    &  1.462$\pm$  0.027    & $-$0.162$\pm$  0.006    &  0.000    &  0.198   \\
  (Ks$-[8.0]$)o    &      BC(Ks)    &  2.781$\pm$  0.034    &  0.506$\pm$  0.028    & $-$0.141$\pm$  0.005    &  0.000    &  0.201   \\
  (Ks$-[8.0]$)o    & BC($[8.0]$)    &  2.781$\pm$  0.034    &  1.506$\pm$  0.028    & $-$0.141$\pm$  0.005    &  0.000    &  0.201   \\

\hline
\end{tabular}
\begin{list}{}
  \item  {\bf Note:} Y-Fit = Coef~n1 + Coef~n2 $\times$ X-axis + Coef~n3 $\times$ X-axis$^2$. \\
  The colours here used are those dereddened with the \Akint\ from surrounding stars
  by \citet{messineo05} and \citet{messineo18}
    (rescaled to a power law with an index of $-2.1$) and the Gordon's extinction curve.
 \item X-ranges are as in Table B.1.
\end{list}
\end{table*}

\begin{figure*}
\begin{center}
\resizebox{0.8\hsize}{!}{\includegraphics[angle=0]{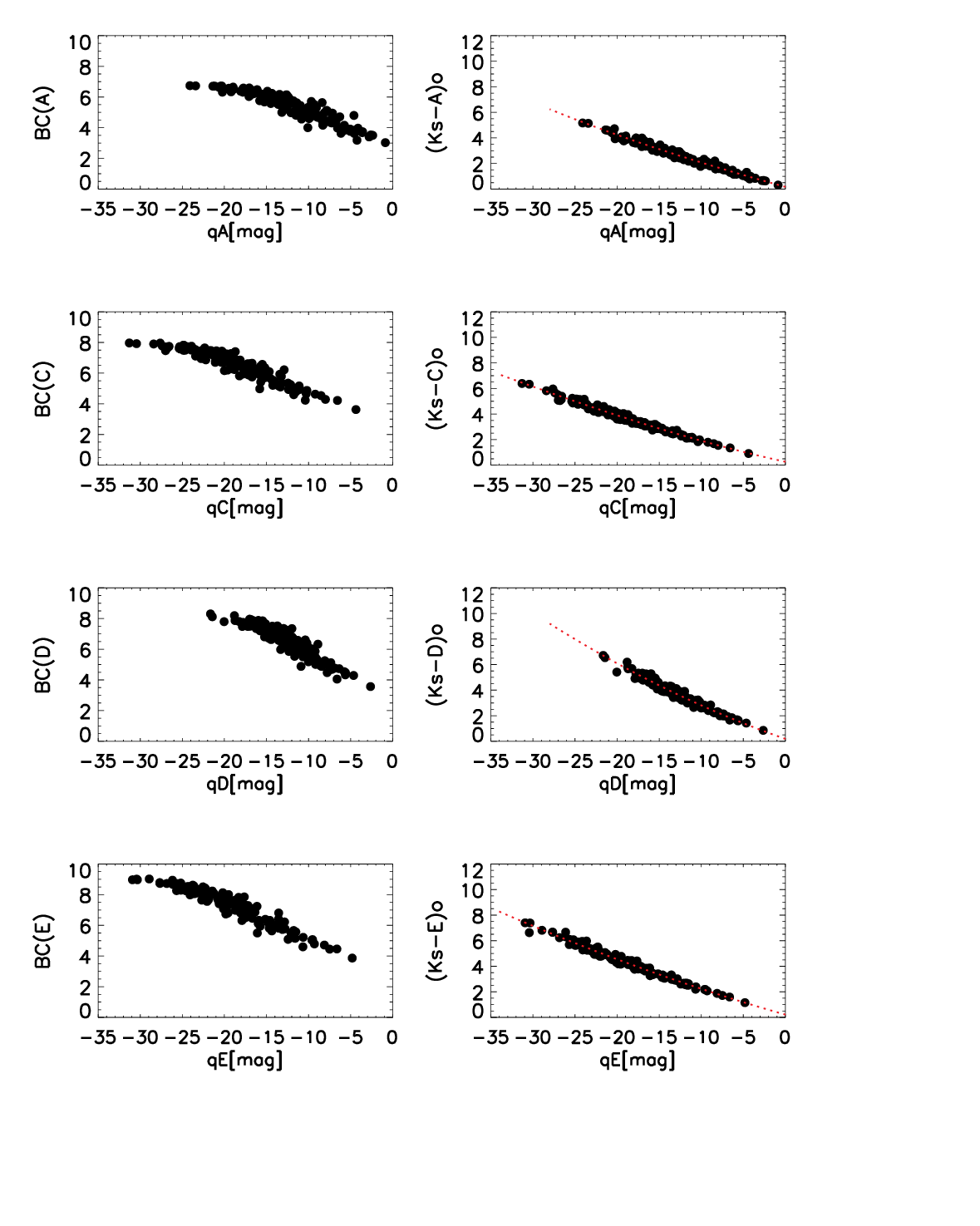}}
\end{center}
\caption{\label{qxmsx} 
Correlations between the   \Ql\ values and the \BCl\ and \Ks$-[\lambda]$ values.
{\it Left panel:} 
\BCl\ values vs. \Ql, for the MSX bands (A, C, D, E).  
{\it Right panel:} De-reddened \Ks$-[\lambda]$ colours  vs. \Ql.
The red dotted lines  are the
 fits given in  Table \ref{table.qxfits}.
 Stars with  MSX A,C,D, and E magnitudes
 available are plotted. }
\end{figure*}

\begin{figure*}
\begin{center}
\resizebox{0.8\hsize}{!}{\includegraphics[angle=0]{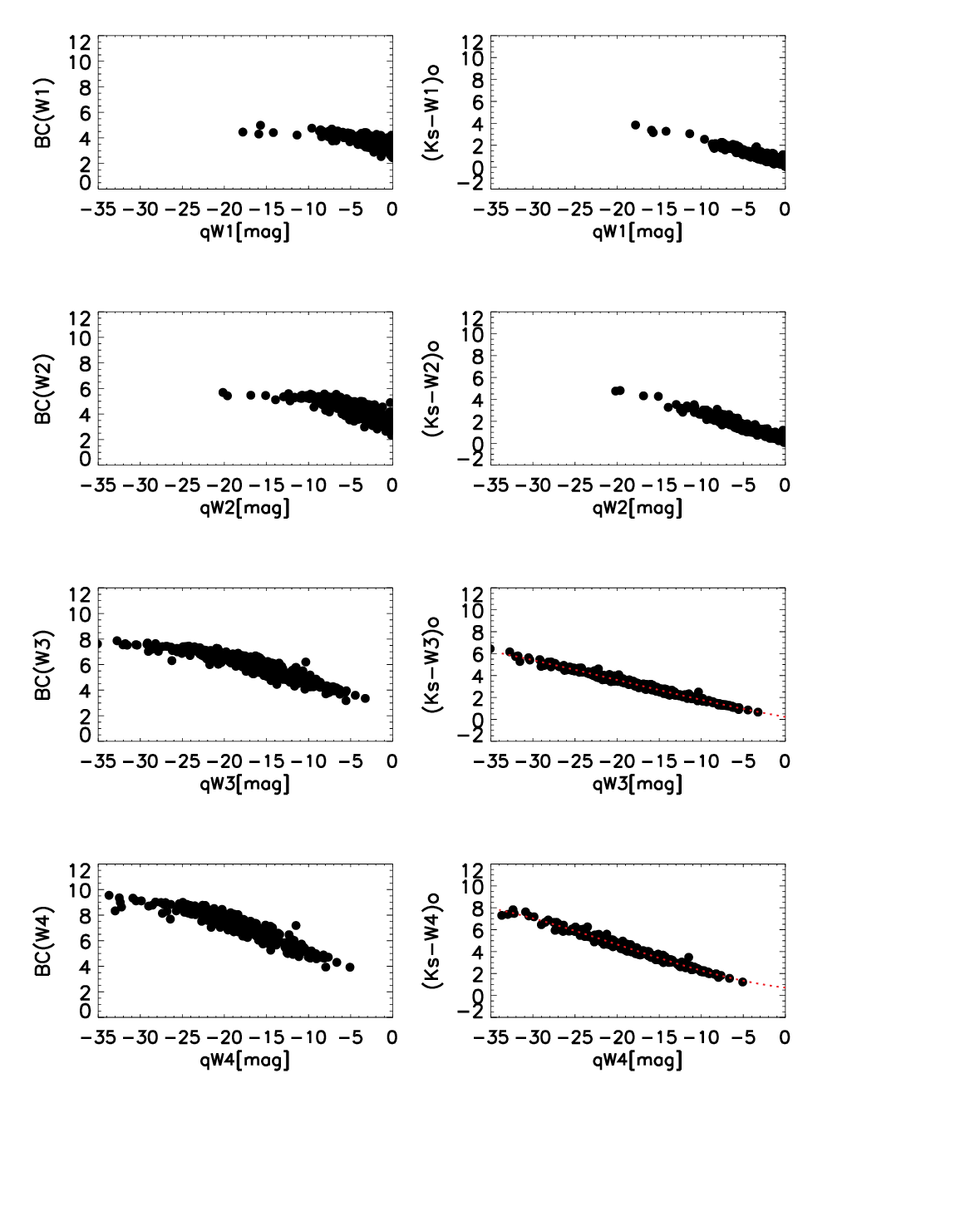}}
\end{center}
\caption{\label{qxwise} \BCl\ values and \Ks$-[\lambda]$ colours vs. \Ql, 
for the WISE bands (W1, W2, W3, and W4). 
Stars with  all four WISE  magnitudes
 available are plotted. The red dotted lines  are the
 fits given in  Table \ref{table.qxfits}.
}
\end{figure*}

\begin{figure*}
\begin{center}
\resizebox{0.8\hsize}{!}{\includegraphics[angle=0]{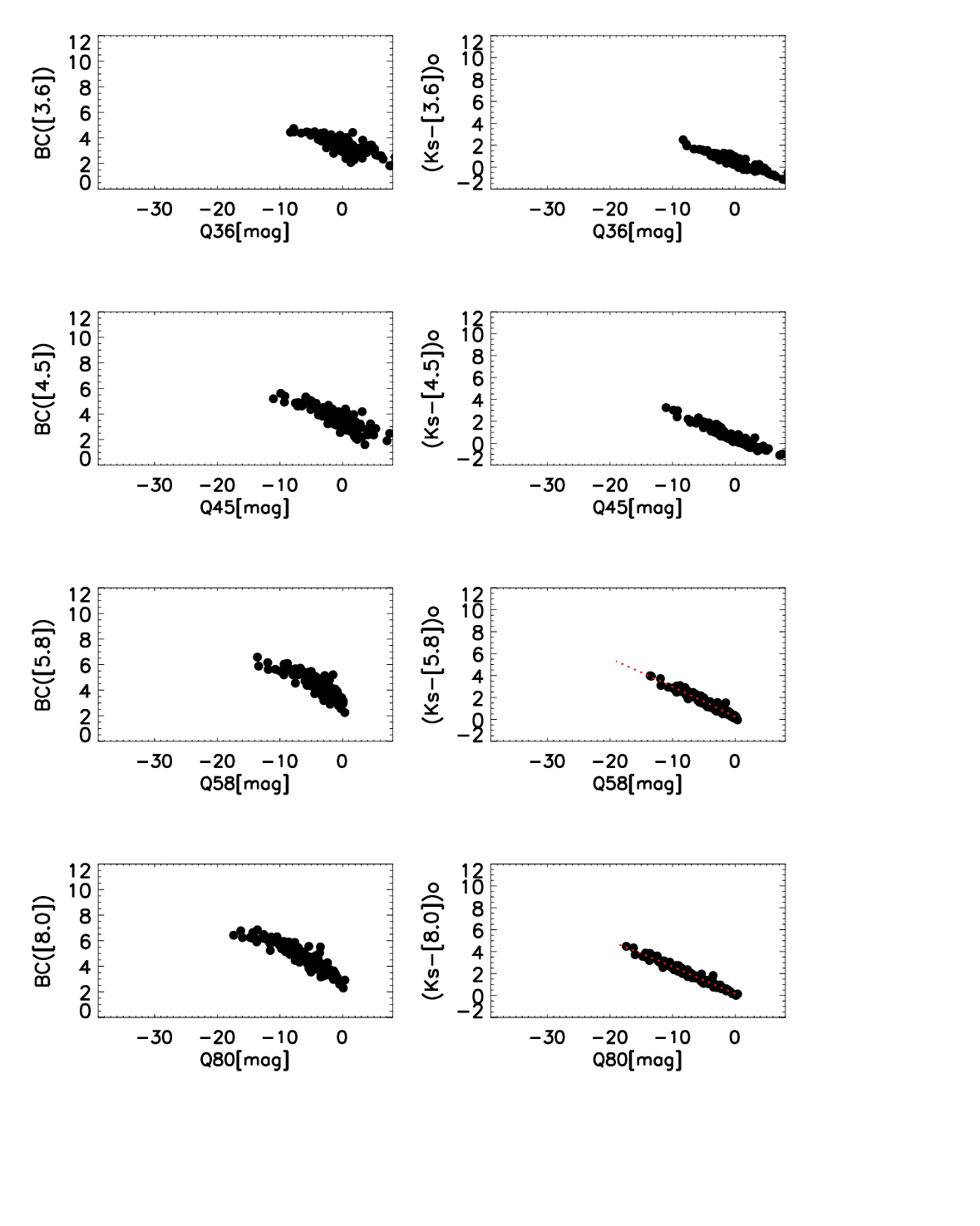}}
\end{center}
\caption{\label{qxglimpse} \BCl\ values and \Ks$-[\lambda]$ colours vs. \Ql, 
for the GLIMPSE bands ($[3.6]$, $[4.5]$, $[5.8]$, and $[8.0]$). 
Stars with  all four GLIMPSE  magnitudes
 available are plotted. The red dotted lines  are the
 fits given in  Table \ref{table.qxfits}.
}
\end{figure*}

The \BCKs\ values and the intrinsic colours are plotted versus the
interstellar-extinction-free parameter \Ql\
in Figs.\ \ref{qxmsx}, \ref{qxwise},
and \ref{qxglimpse} for the MSX, WISE, and GLIMPSE filters.

\end{appendix}

\end{document}